------------------------------------------------
\documentclass[12pt,a4paper]{report}
\usepackage{latexsym}
\begin{document}
\def\beq{\begin{equation}}
\def\eeq{\end{equation}}
\def\bea{\begin{eqnarray}}
\def\eea{\end{eqnarray}}
\def\Real{\mbox{R\hspace{-.9em}I}\,\ }
\newtheorem{theorem}{Theorem}[chapter]
\def\A#1#2#3{\mbox{${#1}_{#2}{}^{#3}$}}
\def\F#1#2{\A{\cal F}{#1}{#2}}
\def\Fe#1#2{\A{F}{#1}{#2}}
\def\f#1#2{\A{f}{#1}{#2}}
\def\R#1#2{\A{R}{#1}{#2}}
\def\T#1#2{\A{T}{#1}{#2}}

\begin{titlepage}
\begin{flushright} 
ITP-UH-3/96  hep-th/9602163
\end{flushright}
\vspace{3cm}
\begin{center} 
{\Large BRS Symmetry and Cohomology}\\[1cm]
Norbert Dragon\\Institut f\"ur Theoretische Physik\\
Universit\"at Hannover\\[3cm]
\end{center}
\noindent
Abstract: The BRS symmetry determines physical states, Lagrange
densities and candidate anomalies. It renders gaugefixing unobservable
in physical states and is required if negative norm states
are to decouple also in interacting models. The relevant mathematical
structures and the elementary cohomological investigations are presented.
\end{titlepage}

This paper is a slightly enlarged version of the lectures given at the
Saalburg Summer School on "Grundlagen und neue Methoden der 
Theoretischen Physik" 1995. It is meant to
give a self contained introduction into one point of view on the 
subject. In particular the mathematical structure is derived completely 
with the exception of the cohomology of simple Lie algebras and
the covariant Poincar' Lemma which are quoted from the literature.

The first chapter deals with the ``raison d'etre'' of gauge symmetries: the
problem to define the subspace of physical states in a Lorentz invariant
theory with higher spin. The operator $Q_s$ which characterizes the physical
states was found by Becchi, Rouet and Stora as symmetry generator of a
fermionic symmetry, the BRS symmetry, in gauge theories with covariant gauge
fixing \cite{brs}. For a derivation of the BRS symmetry from the gauge
fixing in path integrals the reader should consult \cite{beau} or the 
literature quoted there. The chapter is supplemented by a discussion of 
free vector fields for gauge parameter $\lambda \ne 1$. This is not a 
completely trivial exercise \cite{itzub} and not  discussed properly
\cite{henteit} in standard references on gauge systems.

The second chapter deals with the requirement that the physical subspace 
remains physical if interactions are switched on restricts the action to
be BRS invariant. Consequently the Lagrange density has to satisfy a 
cohomological equation similar to the physical states. Quantum 
corrections may violate the requirement of BRS symmetry because
the naive evaluation of Feynman diagrams leads to divergent loop 
integrals which have to be regularized. This regularization can lead to 
an anomalous symmetry breaking which has to satisfy the celebrated
Wess Zumino consistency condition \cite{wesszumino} which again is
a cohomological equation.

In chapter 3 we study some elementary cohomological problems of a 
nilpotent fermionic derivative $d$.
$$
d\omega = 0\quad \omega \bmod d\eta
$$
We derive 
the Poincar\'e Lemma as the Basic Lemma of all the investigations to come.
However, one has to realize that Lagrange densities are defined as 
functions of the fields and their derivatives and not of coordinates. We
investigate differential forms depending on such variables and derive 
the Algebraic Poincar\'e Lemma. The relative cohomology, which 
characterizes Lagrange densities and candidate anomalies, is shown to 
lead to the descent equations which can again be written compactly as a 
cohomological problem. The chapter concludes with K\"unneth's formula 
which allows to tackle cohomological problems in smaller bits if the 
complete problem factorizes.

Chapter 4 is a streamlined version of Brandt's
formulation \cite{brandt1} of the gravitational BRS transformations. In 
this formulation the cohomology factorizes and one has to deal
only with tensors and undifferentiated ghosts. It is shown that the 
ghosts which correspond to translations never occur in anomalies, i.e.
coordinate transformations are not anomalous.

In Chapter 5 we solve the cohomology of the BRS transformations acting 
on ghosts and tensors. The tensors have to couple together with the 
translation ghosts to invariants and also the ghosts for spin and
isospin transformations have to couple to invariants. The invariant 
ghost polynomials generate the Lie algebra cohomology which we quote 
from the mathematical literature \cite{greub3}.  Moreover the tensors 
are restricted by the covariant Poincar\'e Lemma \cite{kovPoinc}. This 
lemma introduces the Chern forms which are the BRS transformation of the
Chern Simons polynomials. 

Chern Simons polynomials and Chern polynomials
are the building blocks of the Chern Simons actions in odd dimensions,
of topological densities and of the chiral anomalies. They are the subject of the 
last chapter. We conclude by giving some well known examples of Lagrange 
densities and anomaly candidates.

The mathematical structures presented in this paper should enable
the reader also to understand and participate in the investigation of 
the master equation which is a still developing field of research
\cite{bbh}. In particular the master equation contains the BRS 
structures for closed algebras but applies also to open algebras.


\chapter{The Space of Physical States}
BRS symmetry is indispensable  in Lorentz covariant
theories with fields with higher spin because it allows to
construct an acceptable space of physical states out of the Fock space 
which contains states with negative norm.

To demonstrate the problem consider the simple example of a 
massless vectorfield
$A_m$. The action $W$ of the vectorfield $A_m, \ m=0,1,2,3$ is
\beq
W[A] = \int d^4x {\cal L}(A(x),\partial A(x))
\eeq
\beq
{\cal L}(A,\partial A) = - \frac{1}{4e^2} (\partial_m A_n - \partial_n A_m)(\partial^m A^n - \partial^n A^m)
-\frac{\lambda}{2e^2}(\partial_m A^m)^2\ .
\eeq
To avoid technical complications at this stage we consider 
the case $\lambda = 1$. The general case is discussed at the end of this
chapter.
We choose to introduce the gauge coupling $e$ here as normalization of the
gauge kinetic energies.
The equations of motion
\beq \frac{\delta W}{\delta A^n(x)} = 
\frac{1}{e^2}\Box A_n(x) 
= 0 \quad \Box=\eta^{mn}\partial_m\partial_n=\partial_0{}^2
-\vec{\partial}^2
\eeq
are solved by the free fields
\bea
\label{free}
A_n(x) = e &\int \frac{d^3k}{(2\pi)^3 2k^0}
(e^{ikx}
a_n^\dagger(\vec{k})
 +
e^{-ikx}
a_n(\vec{k})
)_{\displaystyle |_{k^0 = \sqrt{\vec{k}^2}}}\ .
\eea
They are quantized by the requirement that the propagator
\beq
\label{prop}
\langle 0 | T A_m(x)A_n(0) | 0\rangle
\eeq
be the Greens function of the Euler Lagrange equation
\beq
\label{green}
\frac{1}{e^2}\delta_k^m \Box_x 
\langle 0 | T A_m(x)A^n(0) | 0 \rangle =
i \delta^4(x)\delta_k^n\ .
\eeq
The creation and annihilation operators
$a^\dagger(\vec{k})$  and $a(\vec{k})$ are identified by their commutation
relations with the momentum operators $P^m$
\beq
\left[ P_m , a^\dagger_n(\vec{k})\right] = k_m a^\dagger_n(\vec{k})\quad
\left[ P_m , a_n(\vec{k})\right] = - k_m a_n(\vec{k})
\eeq
which follow because by definition $P_m$ generate translations
\beq
\label{heisen}
\left[i P_m , A_n(x)\right] = \partial_m A_n(x)\ .
\eeq
$a^\dagger_n(\vec{k})$ adds and $a_n(\vec{k})$ subtracts energy 
$k_0=\sqrt{\vec{k}^2}>0$.
Consequently the annihilation operators annihilate the lowest energy state
$|0\rangle$ and justify their denomination
\[P_m |0\rangle = 0 \quad a(\vec{k}) |0\rangle = 0\ . \]
For $x^0 > 0$ the propagator (\ref{prop}) contains only positive frequencies
$e^{-ikx}a_m(\vec{k})$, for $x^0 < 0$ only negative frequencies
$e^{ikx}a^\dagger _m(\vec{k})$. These boundary conditions fix the solution to
(\ref{green}) to be
\beq
\langle 0 | T A_m(x)A_n(0) | 0\rangle =
\lim_{\epsilon\rightarrow 0+}  -i\,e^2\ \eta_{mn}
\int \frac{d^4p}{(2\pi)^4}\frac{e^{ipx}}{p^2 + i\epsilon}\ .
\eeq
Evaluating the $p^0$ integral for positive and for negative $x^0$ and
comparing with the explicit expression for the propagator (\ref{prop}) which
results if one inputs the free fields (\ref{free}) one can read off
$\langle a_m(\vec{k}) a^\dagger_n(\vec{k^\prime})\rangle$ and the value
of the commutator
\beq
\label{comm}
\left[a_m(\vec{k}),a^\dagger_n(\vec{k^\prime})\right]
= -\ \eta_{mn}(2\pi)^3 2 k^0 \delta^3 (\vec{k}-\vec{k^\prime})\ .
\eeq
It is inevitable that the Lorentz metric $\eta_{mn}$ = diag(1,-1,-1,-1)
appears in such commutation relations in Lorentz covariant theories
with fields with higher spin.  The Fock space which results from
such commutation relations necessarily contains negative
norm states because the Lorentz-metric is indefinite and contains both signs.
Consider more specifically the state $|f_0\rangle$
\beq
|f_0\rangle =
\int \frac{d^3k}{(2\pi)^3 2 k^0}f(\vec{k})a^\dagger_0(\vec{k})\ |0\rangle\ .
\eeq
It has negative norm
\[
\langle f_0 | f_0\rangle
= - \eta_{00} \int\frac{d^3k}{(2\pi)^3 2 k^0}|f(\vec{k})|^2 < 0 \ .
\]

In classical electrodynamics (in the vacuum) one does not have the
troublesome amplitude $a^\dagger_0(\vec{k})$. There the wave equation
$\Box A_n =0$ results from Maxwell's equation
$\partial^m(\partial_m A_n - \partial_n A_m) = 0$ and the Lorentz condition
$\partial_m A^m = 0$. This gauge condition fixes the vectorfield up to
the gauge transformation $A_m\rightarrow A^\prime_m = A_m + \partial_m C$
where $C(x)$ satisfies the wave equation $\Box C = 0$.
In terms of the free fields $A_m(x)$ and $C(x)$
\beq
\label{freec}
C(x)= e \int\frac{d^3k}{(2\pi)^3 2 k^0}
\left(e^{ikx}c^\dagger(\vec{k}) +
e^{-ikx}c(\vec{k})\right)_{\displaystyle |_{k^0 = \sqrt{\vec{k}^2}}}
\eeq
one calculates
\[
\partial_m A^m =
i\,e \int\frac{d^3k}{(2\pi)^3 2 k^0}
\left(e^{ikx}k^m a_m^\dagger(\vec{k}) -
e^{-ikx}k^m a_m(\vec{k})\right)_{\displaystyle |_{k^0 = \sqrt{\vec{k}^2}}}
\]
and
\beq
\label{gauge}
A^\prime_m - A_m = \partial_m C =
i\,e \int\frac{d^3k}{(2\pi)^3 2 k^0}
\left(e^{ikx}k_m c^\dagger(\vec{k}) -
e^{-ikx}k_m c(\vec{k})\right)_{\displaystyle |_{k^0 = \sqrt{\vec{k}^2}}}\ .
\eeq
Let us decompose the creation operator $a_m^\dagger(\vec{k})$ into parts in
the direction of the lightlike momentum $k$, in the direction $\bar{k}$ 
(which is $k$ with reflected 3-momentum)
\beq
\bar{k}^m=(k^0,-k^1,-k^2,-k^3)\,
\eeq
and in the two directions $n^i\: i=1,2$ which are orthogonal to $k$ and
$\bar{k}$.
\beq
\label{ahel}
a_m^\dagger(\vec{k})=
\sum_{\tau= k,\bar{k},1,2}
\epsilon_m^{\phantom{m}\tau} a_\tau^\dagger(\vec{k})\ .
\eeq
Explicitly we use polarization vectors $\epsilon^\tau$
\beq
\label{pola}
\epsilon_m^{\phantom{m}\tau} = \left(
\frac{1}{\sqrt2}\frac{k_m}{|\vec{k}|},
\frac{1}{\sqrt2}\frac{\bar{k}_m}{|\vec{k}|},
n_m^1,n_m^2\right)\ \tau = k,\bar{k},1,2
\eeq
with scalar products
\beq
\epsilon^\tau \cdot \epsilon^{\tau^\prime} =
\left(
\begin{array}{llll}
0&1&&\\
1&0&&\\
&&-1&\\
&&&-1
\end{array}\right)\ .
\eeq

The field $\partial_m A^m$ contains the amplitudes
$a^\dagger_{\bar{k}},\ a_{\bar{k}}$.
The Lorentz gauge condition $\partial_m A^m = 0$ eliminates these
amplitudes.

The fields $A^\prime_m$ and $A_m$ differ in the amplitudes
$a^\dagger_{k},\ a_{k}$ in the direction of the momentum $k$.
An appropriate choice of the remaining gauge transformation (\ref{gauge})
cancels these amplitudes.

So in classical electrodynamics $a^\dagger_m$ can be restricted to 2 degrees
of freedom, the transverse oscillations
\[
a_m^\dagger(\vec{k})=
\sum_{\tau= 1,2}
\epsilon_m^{\phantom{m}\tau} a_\tau^\dagger(\vec{k})\ .
\]

The corresponding quantized modes generate a positive definite Fock space.
We cannot, however, just require $a^\dagger_k=0$ and
$a^\dagger_{\bar{k}}=0$ in the quantized theory, this would contradict
the commutation relation
\beq
\left[a_k(\vec{k}),a^\dagger_{\bar{k}}(\vec{k^\prime})\right]
= -\ (2\pi)^3 2 k^0 \delta^3 (\vec{k}-\vec{k^\prime})\ ,
\eeq
which does not vanish. To get rid of the troublesome modes we
require, rather, that physical states do not
contain $a^\dagger_k$ and $a^\dagger_{\bar{k}}$ modes. A slight reformulation
of this condition for physical states leads to BRS symmetry.

To single out a physical subspace of Fock space $\cal F$ we require that
there exists a hermitean operator, the BRS operator,
\beq
Q_s=Q^\dagger_s
\eeq
which defines a subspace $\cal N\subset F$, the gauge invariant states, by
\beq
{\cal N} = \left\{ |\Psi\rangle : Q_s|\Psi\rangle = 0\right\}
\eeq
This requirement is no restriction at all, each subspace can be
characterized as kernel of some hermitean operator.

Inspired by gauge transformations (\ref{gauge}) we take the operator
$Q_s$ to act on one particle states according to
\beq
\label{sa}
Q_s\ a_m^\dagger(\vec{k})|0\rangle = k_m c^\dagger(\vec{k})|0\rangle\ .
\eeq
As a consequence the one particle states generated by
$a_\tau^\dagger(\vec{k})\  \tau = \bar{k},1,2$ belong to $\cal N$.
\beq
Q_s\ a_\tau^\dagger(\vec{k})|0\rangle = 0 \quad \tau = \bar{k},1,2
\eeq
The states created by the creation operator
$a^\dagger_k$ in the direction of the momentum $k$ are not invariant
\[
Q_s\ a_k^\dagger(\vec{k})|0\rangle =
\sqrt{2} |\vec{k}| c^\dagger(\vec{k})|0\rangle\ \ne 0
\]
and do not belong to $\cal N$.

The space $\cal N$ is not yet acceptable because is contains
non vanishing zero-norm states
\beq
\label{zero}
|f\rangle =
\int \tilde{d}k f(\vec{k}) a_{\bar{k}}^\dagger(\vec{k})|0\rangle\quad
\langle f | f \rangle = 0\ {\rm  because }\
\left[a_{\bar{k}}(\vec{k}),a^\dagger_{\bar{k}}(\vec{k^\prime})\right]
= 0\ .
\eeq
To get rid of these states the following observation is crucial:
\begin{theorem}{\quad}\\
\label{t1}
Scalar products of gauge invariant states 
$|\psi\rangle \in \cal N$ and
$|\chi\rangle \in \cal N$
remain unchanged if the state $|\psi\rangle$ is replaced by
$|\psi\rangle + Q_s |\Lambda\rangle $ .
\end{theorem}
Proof:
\beq
\label{scal}
\langle\chi|\left( |\psi\rangle + Q_s |\Lambda\rangle \right) =
\langle\chi|\psi\rangle + \langle\chi|Q_s |\Lambda\rangle =
\langle\chi|\psi\rangle
\eeq
The term $\langle\chi|Q_s |\Lambda\rangle$ vanishes, because $Q_s$ is
hermitean and $Q_s|\chi\rangle = 0$ .

We arrive at the BRS algebra from the seemingly innocent requirement that
$|\psi\rangle + Q_s |\Lambda\rangle$ belongs to $\cal N$ whenever
$|\psi\rangle $ does. The requirement seems natural because
$|\psi\rangle + Q_s |\Lambda\rangle$ and $|\psi\rangle $ have the same
scalar products with gauge invariant states and therefore cannot be
distinguished experimentally. It is, nevertheless, a most restrictive condition,
because it requires $Q_s^2$ to vanish on each state $|\Lambda\rangle$, i.e.
$Q_s$ is nilpotent.
\beq
Q_s^2 = 0
\eeq
We require this relation as defining property of the BRS operator.
Then the space $\cal N$ of gauge invariant states decomposes into equivalence
classes
\beq
\label{equiv}
|\psi\rangle \sim |\psi\rangle + Q_s |\Lambda\rangle\ .
\eeq
These equivalence classes are the physical states.
\beq
\label{phys}
{\cal H}_{phys}=\frac{\cal N}{Q_s \cal F} =
\left\{ |\psi\rangle : Q_s |\psi\rangle = 0 \ , |\psi\rangle\  {\rm mod}\  Q_s|\Lambda\rangle
\right\}
\eeq
${\cal H}_{phys}$ inherits a scalar product from $\cal F$ because the scalar
product in $\cal N$ does not depend on the representative of the equivalence
class by theorem \ref{t1}.

The construction of ${\cal H}_{phys}$ by itself does not guarantee that
${\cal H}_{phys}$ has a positive definite scalar product. This will hold
only if $\cal F$ and $Q_s$ are suitably chosen. One has to check this
positive definiteness in each model.

In the case at hand, the zero-norm states $|f\rangle$ (\ref{zero})
 are equivalent to 0
in ${\cal H}_{phys}$ if there exists a massless, real field $\bar{C}(x)$
\beq
\label{cq}
\bar{C}(x)= e \int\frac{d^3k}{(2\pi)^3 2 k^0}
\left(e^{ikx}\bar{c}^\dagger(\vec{k}) +
e^{-ikx}\bar{c}(\vec{k})\right)_{\displaystyle |_{k^0 = \sqrt{\vec{k}^2}}}
\eeq
and if $Q_s$ transforms the one-particle states according to
\beq
\label{scq}
Q_s \bar{c}^\dagger(\vec{k}) |0\rangle =
i \sqrt{2} |\vec{k}| a^\dagger_{\bar{k}}(\vec{k})|0\rangle\ .
\eeq
For the six one-particle states we conclude that
$\bar{c}^\dagger(\vec{k}) |0\rangle$ and
$a^\dagger_{k}(\vec{k}) |0\rangle$ are not invariant (not in $\cal N$),
$a^\dagger_{\bar{k}}(\vec{k}) |0\rangle$  and
$c^\dagger(\vec{k}) |0\rangle$ are of the form $Q_s |\Lambda\rangle$ and
equivalent to 0, the remaining two transverse creation operators
generate the physical one particle space with positive norm.

Notice the following pattern: states from the Fock space $\cal F$ are
excluded in pairs from the physical Hilbert space ${\cal H}_{phys}$,
one state $|n\rangle$ is not invariant 
\beq
\label{nt}
Q_s|n\rangle = |t\rangle \ne 0
\eeq 
and therefore not contained in $\cal N$, the other $|t\rangle$ is trivial
and equivalent to 0 in ${\cal H}_{phys}$ because it is the transform of
$|n\rangle$ :  $|t\rangle = Q_s |n\rangle$.

The algebra $Q_s^2=0$ enforces 
\beq
\label{t0}
Q_s |t\rangle = 0\ .
\eeq
If one uses $|t\rangle$
and $|n\rangle$ as basis then $Q_s$ is represented by the matrix
\beq
\label{mat2}
Q_s = \left( {0\ 1 \atop 0\ 0}\right)\ .
\eeq
This is one of the two possible Jordan block matrices which can represent
a nilpotent operator $Q_s^2=0$. The only eigenvalue is 0, so a Jordan block
consists of a matrix with zeros and with 1 only in the upper diagonal
\[
Q_{s\,ij}=\delta_{i+1,j}\ .
\]
Because of $Q_s^2 = 0$ the blocks can only have the size
$1\times 1$ or $2\times 2$. In the first case the corresponding vector on
which $Q_s$ acts is invariant and not trivial and contributes to
${\cal H}_{phys}$. The second case is given by (\ref{mat2}), the corresponding
vectors are not physical.

It is instructive to consider the scalar product of the states on which
$Q_s$ acts. If it is  positive definite then $Q_s$ has to vanish because $Q_s$
is hermitean and can be diagonalized in a space with positive definite scalar
product. Thereby the non diagonalizable $2\times2$ block (\ref{mat2}) 
would be excluded. It is,
however, in Fock spaces with indefinite scalar product that we need the
BRS operator and there it can act nontrivially. In the physical Hilbert space,
which has a positive definite scalar product, $Q_s$ vanishes.
Nevertheless the existence of the BRS operator $Q_s$ in Fock space severely
restricts the possible actions of the models we are going to consider.

Reconsider the doublet (\ref{nt}, \ref{t0}): one can
easily verify that by suitable choice of $|n\rangle$ and
$|t\rangle$
the scalar product (if it is non-degenerate) can be brought to one of the two
standard forms
\beq
\langle n|n\rangle = 0 = \langle t|t\rangle\quad
\langle t|n\rangle =  \langle n|t\rangle = 1 {\rm\  or\ } (-1)\ .
\eeq
This is an indefinite scalar product of Lorentzian type
\beq
|e_\pm\rangle= \frac{1}{\sqrt{2}}(| n\rangle \pm | t\rangle) \quad
\langle e_+|e_-\rangle = 0 \quad
\langle e_+|e_+\rangle = - \langle e_-|e_-\rangle = 1\  {\rm or}\ (-1)\ .
\eeq
By our construction (\ref{phys}) of ${\cal H}_{phys}$ pairs of states with
wrong sign norm and with acceptable norm are excluded from the physical
states.

Let us close this chapter with a supplement which describes free vector 
fields for $\lambda \neq 1$. They have to satisfy the equations of 
motion
\beq
\label{eqnln1}
\frac{1}{e^2}(\Box A_n + (\lambda - 1)\partial_n \partial_m A^m )= 0\ .
\eeq
It is easy to derive from this the necessary condition 
\beq
\label{box2}
\Box\Box A_m = 0
\eeq
and its Fourier transformed version 
$$(p^2)^2\tilde{A}_m=0\ .$$
From this
one can conclude that $\tilde{A}$ vanishes outside the 
light cone and that the general solution $\tilde{A}$ contains a $\delta$ 
function and its derivative.
$$
\tilde{A}_m = a(p)\delta(p^2) + b(p)\delta^\prime (p^2)
$$
However, the derivative of the $\delta$ function is ill defined because 
spherical coordinates $p^2,v,\vartheta,\varphi$ are discontinuous
at $p=0$.

To solve $\Box\Box \phi = 0$ one can restrict $\phi(t,\vec{x})$ to
$\phi(t)e^{i\vec{k}\vec{x}}$, the general solution can then be 
obtained as a wavepaket which is superposed out of solutions of this 
form. $\phi(t)$ has to satisfy the 
ordinary differential equation 
$$(\frac{d^2}{dt^2}-k^2)^2\phi=0$$
which  has the general solution $$\phi(t)=(a+bt)e^{i\omega t}\ .$$ 
Therefore the equations (\ref{box2}) are solved by
\bea
A_n(x) &=& \int \frac{d^3k}{(2\pi)^3 2k^0}
\left(e^{ikx} a_n^\dagger(\vec{k}) +
x^0 e^{ikx} b^\dagger_ n(\vec{k})\ + \right. \nonumber \\
&\phantom{=}&\phantom{\int \frac{d^3k}{(2\pi)^3 2}}
\left.
+\ e^{-ikx} a_n(\vec{k})
+x^0 e^{-ikx} b_n(\vec{k})
\right)_{\displaystyle |_{k^0 = \sqrt{\vec{k}^2}}}\ .
\eea
This equation makes the vague notion $\delta^\prime(p^2)$ explicit.
The amplitudes $b_n$, $b^\dagger_n$ are determined from the coupled 
equations (\ref{eqnln1}).
\bea
\label{freeln1}
A_n(x) &=& e\,\int \frac{d^3k}{(2\pi)^3 2k^0}
\left(e^{ikx} a_n^\dagger(\vec{k})
-i\frac{\lambda -1}{\lambda + 1}x^0 e^{ikx} \frac{k_n}{k_0} k^m a^\dagger_ m
(\vec{k})\ + \right. \nonumber \\
&\phantom{=}&\phantom{\int \frac{d^3k}{(2\pi)^3 2}}
\left.
+\ e^{-ikx} a_n(\vec{k})
+i\frac{\lambda -1}{\lambda + 1}x^0 e^{-ikx} \frac{k_n}{k_0} k^m a_m(\vec{k})
\right)_{\displaystyle |_{k^0 = \sqrt{\vec{k}^2}}}
\eea
From (\ref{heisen}) one can deduce that the commutation relations
\beq
[P^i,a^\dagger_m(\vec{k})]=k^ia^\dagger_m(\vec{k})\quad 
[P^i,a_m(\vec{k})]=-k^ia_m(\vec{k})\quad i=1,2,3
\eeq
and 
\beq
\label{hacom}
[P_0,a^\dagger_m(\vec{k})]=k_0a^\dagger_m(\vec{k})
-\frac{(\lambda-1)}{(\lambda +1)}\frac{k_m}{k_0}k^na^\dagger_n(\vec{k})
\eeq
have to hold.
If we decompose $a^\dagger_m(\vec{k})$ according to (\ref{ahel}) then we
obtain
\beq
[P_0,a^\dagger_t(\vec{k})]=k_0a^\dagger_t(\vec{k}) \quad t=1,2
\eeq
for the transverse creation operators and also
\beq
[P_0,a^\dagger_{\bar{k}}(\vec{k})]=k_0a^\dagger_{\bar{k}}(\vec{k}) 
\eeq
for the creation operator in direction of $\bar{k}$. For the creation
operator in the direction of the four momentum $k$ one gets
\beq
[P_0,a^\dagger_k(\vec{k})]=k_0a^\dagger_k(\vec{k})
-2k_0\frac{\lambda-1}{\lambda +1}a^\dagger_{\bar{k}}(\vec{k})\ .
\eeq
In particular, for $\lambda \neq 1$, $a^\dagger_k(\vec{k})$ does not 
generate energy eigenstates and the hermitean operator $P_0$ cannot be
diagonalized in Fock space because the commutation relations are given 
by
$$
[P_0,a^\dagger]=M\,a^\dagger
$$
with a matrix $M$ which contains a nondiagonalizable Jordan block
\beq
M\sim k_0\left( 
\begin{array}{c c}
1&-2\frac{\lambda-1}{\lambda+1}\\
0&1
\end{array}
\right )
\eeq 
That hermitean operators are not guaranteed to be diagonalizable is of 
course related to the indefinite norm in Fock space. For operators 
$O_{phys}$ which correspond to measuring devices it is sufficient that
they can be diagonalized in the physical Hilbert space. This is 
guaranteed if ${\cal H}_{phys}$ has positive norm. In Fock space it is 
sufficient that operators $O_{phys}$  commute with the BRS operator
$Q_s$ and that they satisfy generalized eigenvalue equations
\beq
O_{phys}|\psi_{phys}\rangle = c |\psi_{phys}\rangle +
 Q_s |\chi\rangle \quad c \in \Real
\eeq
from which the spectrum can be read off.

The Hamilton operator $H=P_0$ which results from the Lagrange density
\beq
{\cal 
L}=-\frac{1}{4e^2}F_{mn}F^{mn}-\frac{\lambda}{2e^2}(\partial_mA^m)^2 
\eeq
\beq
H=\frac{1}{2e^2}\!\int \!d^3x :\!\!\left( (\partial_0 A_i)^2 
-(\partial_iA_0)^2
- \lambda(\partial_0A_0)^2+(\partial_jA_i-\partial_iA_j)^2 +
\lambda(\partial_iA_i)^2\right )\!\! :
\eeq
can be expressed in terms of the creation and 
annihilation operators.
\beq
H=\int \frac{d^3k}{(2\pi)^3 2k_0}k_0\left (
\sum_{t=1}^{2}a^\dagger_ta_t  - \frac{2\lambda}{\lambda +1}
\left(a^\dagger_ka_{\bar{k}}+a^\dagger_{\bar{k}}a_k-2\frac{\lambda-1}{
\lambda + 1} a^\dagger_{\bar{k}}a_{\bar{k}}\right) \right)
\eeq

$H$ satisfies (\ref{hacom}) because the creation and annihilation
operators fulfil  the commutation relations
\beq
\label{creannco}
[a_m(\vec{k}),a^\dagger_n(\vec{k}^\prime)]=2k^0(2\pi)^3\delta^3(\vec{k}
- \vec{k}^\prime)\!\left (-\eta_{mn}+\frac{\lambda -1}{2\lambda k^0}
(\eta_{m0}k_n + \eta_{n0}k_m - k_m k_n)\right )
\eeq
which follow from the requirement that the propagator
\beq
\langle {\rm T}A_m(x)A^n(0)\rangle = -i e^2 \lim_{\varepsilon 
\rightarrow 0\scriptscriptstyle{+}}\int 
\frac{d^4p}{(2\pi)^4}\frac{e^{{ipx}}}{(p^2+i\varepsilon)^2}
\left(p^2\delta_m{}^n-\frac{\lambda-1}{\lambda}p_mp^n\right )
\eeq
is the Greens function corresponding to the equation of motion 
(\ref{eqnln1}).
If one decomposes the creation annihilation operators according to 
(\ref{ahel}) then the transverse operators satisfy
\beq
[a_i(\vec{k}),a^\dagger_j(\vec{k}^\prime)]=2k^0(2\pi)^3\delta^3(\vec{k}
- \vec{k}^\prime)\delta_{ij}\quad i,j \in \{1,2\}
\eeq
They commute with the other creation annihilation operators which have 
the following off diagonal commutation relations
\beq                                                                              
[a_{\bar{k}}(\vec{k}),a^\dagger_k(\vec{k}^\prime)]
=[a_{{k}}(\vec{k}),a^\dagger_{\bar{k}}(\vec{k}^\prime)]
=-\frac{\lambda +1}{2\lambda}2k^0(2\pi)^3\delta^3(\vec{k}- 
\vec{k}^\prime)\ .
\eeq
The other commutators vanish.

The analysis of the BRS transformations leads again to
the result that physical states are generated only by the transverse
creation operators.


\chapter{BRS symmetry}
To choose the physical states one could have proceeded like
Cinderella and pick acceptable states by hand or have them picked by
doves. Prescribing the action of $Q_s$ on one particle states (\ref{sa},
\ref{scq}) is not really different from such an arbitrary approach.
From (\ref{sa},\ref{scq}) we know nothing about physical multiparticle
states. Moreover we would like to know whether one can switch on interactions
which respect our definition of physical states. Interactions should
give transition amplitudes which are independent of the choice (\ref{equiv})
of the representative of physical states. The time evolution should
leave physical states physical, otherwise negative norm states could result
from physical initial states.

All these requirements can be satisfied if the BRS  operator $Q_s$ 
belongs
to a symmetry. We interpret the equation $Q_s^2=0$ as a graded commutator,
an anticommutator, of a fermionic generator of a Lie algebra
\beq
\label{qsq}
\{Q_s,Q_s\} = 0\ .
\eeq

To require that $Q_s$ be fermionic means that the BRS  operator 
transforms fermionic variables into bosonic variables and vice versa. In
particular we take $A_m(x)$ to be a bosonic field. Then the fields
$C(x)$ and $\bar{C}(x)$ have to be fermionic
though they are real scalar fields and carry no spin. They violate the
spin  statistics relation which requires physical fields with half-integer
spin to be fermionic and fields with integer spin to be bosonic. This
violation can be tolerated because the corresponding particles do not
occur in physical states, they are ghosts. We call $C(x)$ the ghost field and
$\bar{C}(x)$ the antighost field. Because the ghost fields $C$ and $\bar{C}$
anticommute they contribute, after introduction of interactions, to loop
corrections with the opposite sign as compared to bosonic contributions.
The ghosts compensate in loops for the unphysical bosonic degrees of freedom
contained in the field $A_m(x)$.

We want to realize the algebra (\ref{qsq}) as local transformations
on fields. Then we have to determine actions which are invariant
under these transformations and construct the BRS  operator as Noether charge
corresponding to this symmetry.

The transformations act on commuting and anticommuting classical variables,
the fields, and polynomials in these fields, the Lagrange densities.
We write the commutation relation as
\beq
\phi^i\phi^j=(-1)^{|\phi^i|\cdot|\phi^j|}\phi^j\phi^i =: (-)^{ij}\phi^j\phi^i
\eeq
Here we have introduced the grading
\beq
|\phi^i|=\left\{ 
\begin{array}{l}
0 {\rm \ if\ }\phi^i{\rm \ is\  bosonic}\\
1 {\rm \ if\ }\phi^i{\rm \ is\  fermionic}
\end{array}
\right.
\eeq
and a shorthand $(-)^{ij}$ for $(-1)^{|\phi^i|\cdot|\phi^j|}$. Because products
are understood to be associative monomials get a natural grading
\beq
|\phi^i\phi^j|=|\phi^i|+|\phi^j| \bmod 2\ .
\eeq
We will consider only polynomials which are sums of monomials with the 
same grading, these polynomials are graded commutative
\beq
A B = (-1)^{|A|\cdot|B|}B A\ .
\eeq
Transformations and symmetries are operations $O$ acting linearly, i.e. term
by term, on polynomials. We consider only operations which map 
polynomials with a definite grading to polynomials with a definite 
grading. These operations have a natural grading.
\bea
O(\lambda_1 A + \lambda_2 B) &=& \lambda_1 O(A) + \lambda_2 O(B)\\
|O|&=&|O(A)| - |A| \bmod 2
\eea
Derivative operators 
\footnote{More precisely this Leibniz rule defines left 
derivatives. The left factor $A$ is differentiated without a 
graded sign}
$v$ of first order satisfy in addition a graded 
Leibniz rule 
\beq
\label{leibniz}
v(AB)=(vA)B + (-)^{|v|\cdot|A|}A(vB)\ .
\eeq
They are completely determined by their action on the elementary variables $\phi^i$:
$v(\phi^i)=v^i$, i.e. $v=v^i\partial_i$. The partial derivatives 
$\partial_i$
are naturally defined by $\partial_i\phi^j=\delta_i^{\phantom{i}j}$.
They have the same grading as their corresponding variables
$|\partial_i| = |\phi^i|$, the
grading of the components $v^i$ results naturally$|v^i|=|v|+|\phi^i|\bmod 2$.

An example of a fermionic derivative is given by the exterior derivative
\beq
\label{dext}
d=dx^m\partial_m
\eeq
It transforms coordinates $x^m$ into differentials $dx^m$ which have 
opposite statistics
\beq
|dx^m|=|x^m| + 1 \bmod 2 
\eeq
and which commute with $\partial_n$
\beq
\quad [\partial_n,dx^m] = 0\ .
\eeq
Therefore the exterior derivative is nilpotent
\beq
d^2=0\ .
\eeq

Lagrange densities have to be real polynomials to make the resulting 
$S$-matrix
unitary. This is why we have to discuss complex conjugation. We
define conjugation
such that hermitean conjugation of a time ordered operator 
corresponding to some polynomial gives the
antitime ordered operator corresponding to the conjugated polynomial. We
therefore require 
\bea
(\lambda_1 A + \lambda_2 B)^\ast &=& 
\lambda_1^\ast A^\ast + \lambda_2^\ast B^\ast \\
(AB)^\ast &=& B^\ast A^\ast = (-)^{|A||B|}A^\ast B^\ast\\
|\phi^\ast| &=& |\phi| \ .
\eea
Conjugation preserves the grading
and is defined on polynomials whenever
it is defined on the elementary variables $\phi^i$. It can be used to define
conjugation of operations $O$ (they map polynomials to polynomials and have
to be distinguished from operators in Fock space).
\beq
\label{opcon}
O^\ast(A)=(-)^{|O||A|}(O(A^\ast))^\ast
\eeq
This definition ensures that $O^\ast$ is linear and satisfies the Leibniz rule
if $O$ is a first order derivative. 

The exterior derivative $d$ is real if the conjugate differentials are 
given by 
\beq
\quad (dx^m)^\ast = (-)^{|x^m|}d((x^m)^\ast)\ .
\eeq
The partial derivative with respect to a real fermionic variable
is purely imaginary as is the operator $\delta$
\beq
\delta=x^m\frac{\partial}{\partial (dx^m)} \quad \delta^\ast = - \delta 
\ . \eeq

The anticommutator $\Delta$
\beq
\Delta=\{d,\delta\}= x^m\frac{\partial}{\partial x^m} + 
dx^m\frac{\partial}{\partial (dx^m)} = N_x + N_{dx}
\eeq
which counts the variables $x$ and $dx$ is again real because the 
definition (\ref{opcon}) implies \beq
(O_1O_2)^\ast=(-)^{|O_1||O_2|}O_1^\ast O_2^\ast\ .
\eeq
Conjugation does \underline{not} change the order of two operations 
$O_1$ and $O_2$.

We can now define the BRS  transformation
$s$. It is a real, fermionic, nilpotent first order derivative.
\beq 
s=s^\ast \quad |s|=1 \quad s^2=0
\eeq
It acts on Lagrange densities and functionals of fields. 
Space-time derivatives $\partial_m$ of fields are limits of differences of
fields taken at neighbouring arguments. It follows from the linearity
of $s$ that it has to commute with space-time derivatives
\beq
\label{sder}
[s,\partial_m]=0\ .
\eeq
Linearity  implies moreover that the BRS  transformation of integrals
is given by the integral of the transformed integrand. Therefore the 
differentials $dx^m$ are BRS  invariant
\beq
s(dx^m) = 0 = \{s,dx^m\} \quad ( [s,dx^m]=0 \mbox{ for fermionic }x^m )
\eeq
Taken together the last two equations imply that $s$ and $d$ 
(\ref{dext}) anticommute 
\beq
\{s,d\}=0
\eeq

In the simplest multiplet $s$ transforms a real anticommuting field 
$\bar{C}(x)=\bar{C^\ast}(x)$, the antighost field, into $\sqrt{-1}$ 
times a real bosonic field
$B(x)=B^\ast(x)$, the auxiliary field. The denominations will be justified
once the Lagrange density is given.
\beq
\label{scqt}
s\bar{C}(x)=iB(x)\quad sB(x)=0
\eeq
The BRS  transformation which corresponds to abelian gauge transformations
acts on a real bosonic vectorfield $A_m(x)$ and a real, fermionic ghost
field $C(x)$ by
\beq
\label{sam}
sA_m(x)=\partial_mC(x) \quad sC(x)=0\ .
\eeq

We can attribute to the fields 
\beq 
\phi=\bar{C},B,A_m,C
\eeq
and to $s$ a ghost
number 
\beq
{\rm gh}(\bar{C})= -1,\  {\rm gh}(B)=0,\  
{\rm gh}(A_m)=0,\  {\rm gh}(C)=1\ .
\eeq
\beq
{\rm gh}(s)=1\ .
\eeq

We anticipate the analysis of the algebra (\ref{scqt}, \ref{sam})
and state the result in $D=4$ dimensions\footnote{The result holds
more generally in even dimensions. In odd dimensions Chern-Simons 
forms can occur in addition}. All invariant, local actions \beq
W[\phi] =\int d^4x {\cal L}(\phi,\partial\phi,\partial\partial\phi,\dots
)
\eeq
with ghostnumber $0$ have the form
\beq
\label{lagr}
{\cal L} = {\cal L}_{inv}(F_{mn},\partial_l F_{mn},\dots) 
+ i s X(\phi,\partial \phi, \dots )\ .
\eeq
The part ${\cal L}_{inv}$ is real, it depends only on the field
strengths
\beq
F_{mn}= \partial_m A_n -\partial_n A_m
\eeq
and their partial derivatives.
Therefore it is invariant under classical gauge transformations. 
Typically it is chosen to be
\beq
\label{linv}
{\cal L}_{inv}=-\frac{1}{4e^2}F_{mn}F^{mn}\ .
\eeq
The gauge coupling constant $e$ is introduced as normalization
of the kinetic energy of the gauge field. 

The function
$X(\phi,\partial\phi,\dots)$ is a real, fermionic polynomial with ghostnumber
gh$(X)=-1$. It has to contain a factor $\bar{C}$ and is in the simplest case
given by 
\beq
X=\frac{\lambda}{e^2}\bar{C}(-\frac{1}{2}B+\partial_m A^m)\ .
\eeq
$\lambda$ is the gauge fixing parameter. The piece $i s X$ contributes the
gaugefixing for the vectorfield and contains the action of the ghostfields
$C$ and $\bar{C}$.
\beq
\label{lgf}
i s X = \frac{\lambda}{2 e^2}(B - \partial_m A^m)^2 
-\frac{\lambda}{2 e^2}(\partial_m A^m)^2 -i\frac{\lambda}{e^2}
\bar{C}\partial_m\partial^mC
\eeq
This Langrange density makes $B$ an auxiliary field, its equation
of motion fix it algebraically  $B = \partial_m A^m$. $C$ and $\bar{C}$
are free fields (\ref{freec}, \ref{cq}).

To justify the name gauge fixing for the gauge breaking part
$-\frac{\lambda}{2 e^2}(\partial_m A^m)^2$ of the Lagrange density
we show that a change of the fermionic function $X$  cannot be measured in
amplitudes of physical states as long as such a change leads only to
a differentiable perturbation of amplitudes. This means that gauge fixing
and ghostparts of the Lagrange density are unobservable. Only the parameters
in the gauge invariant part ${\cal L}_{inv}$ are measurable.
\begin{theorem}{\quad}\\
Transition amplitudes of physical states are independent of the gauge fixing
(within perturbatively connected gauge sectors).
\end{theorem}
Proof: If one changes $X$ by $\delta X$ then the Lagrange density and the action 
change by
\beq
\delta {\cal L} = i\, s\,\delta X \quad \delta W =
i\,s\int d^4x\,\delta X\ .
\eeq
$S$-matrix elements of physical states $|\chi\rangle$ and $|\psi\rangle$
change to first order by
\beq
\delta \langle\chi_{out}|\psi_{in}\rangle = 
\langle\chi_{out}|i\cdot i \int d^4x\,s\, \delta X |\psi_{in}\rangle
\eeq
where $s\,\delta X$ is an operator in Fock space.
The transformation $s\,\delta X$ of the operator $\delta X$
is generated by $i$ times the
anticommutator of the fermionic operator $\delta X$ with the fermionic
BRS  operator $Q_s$
\beq
\langle\chi_{out}| s\int d^4x\, \delta X |\psi_{in}\rangle =
\langle\chi_{out}| [i\,Q_ s,\int d^4x\,\delta X]_+ |\psi_{in}\rangle\ .
\eeq
This expression vanishes because $|\chi\rangle$ and $|\psi\rangle$ are
physical (\ref{phys}) and $Q_s$ is hermitean.

The proof does not exclude the possibility that there exist different sectors
of gauge fixing which can be distinguished and cannot be joined by a
perturbatively smooth change of parameters.

Using this theorem we can concisely express the restriction which the
Lagrange density of a local, BRS  invariant action in $D$ dimensions has 
to satisfy.

It is advantageous to combine $\cal L$ with the
differential $d^Dx$ and consider the Lagrange density as a $D$-form
$\omega^0_{D}= {\cal L}d^Dx$ with ghostnumber 0.
The BRS  transformation of the Lagrange density $\omega^0_{D}$
has to give a (possibly vanishing) total derivative $d\,\omega_{D-1}^1$.

With this notation the condition for an invariant local action is
$$
s\,\omega^0_{D} + d\,\omega^1_{D-1} = 0\ .
$$

It is sufficient to determine this Lagrange density $\omega^0_{D}$ up to
a piece of the form $s\,\eta^{-1}_{D}$, where $\eta^{-1}_{D}$ carries
ghostnumber -1. Such a piece contributes only to gaugefixing
and to the ghostsector and cannot be observed. It is trivially BRS  invariant
because $s$ is nilpotent. A total derivative part $d\,\eta^0_{D-1}$ 
(with gh$(\eta^0_{D-1})=0 $) of the Lagrange
density contributes only boundary terms to the action and is also neglected.
This means that we look for the solutions of the equation
\beq
\label{relcoh}
s\,\omega^0_{D} + d\,\omega^1_{D-1} = 0\quad
\omega^0_{D} \ \bmod\ ( s\eta^{-1}_{D} + d\eta^0_{D-1} )\ .
\eeq

This is a cohomological equation and very similar to the equation which
determines the physical states (\ref{phys}). The Equivalence classes of 
solutions $\omega^0_{D}$ of this equations span a linear space: the
relative cohomology of $s \bmod d$ with ghost number indicated by the
superscript and form degree denoted by the subscript.

If we use a Lagrange density which solves this equation, then the action
is invariant under the continuous symmetry 
$\phi \rightarrow \phi +\, \alpha\, s\ \phi$ with an arbitrary fermionic
parameter $\alpha$. In classical field theory Noether's theorem then
guarantees that there exists a current $j^m$ which is conserved as a
consequence of the equations of motion. The integral 
$Q_s=\int d^3x\,j^0$ is constant in time and generates
the nilpotent BRS  transformations
\beq
s A = \{ A,Q_s\}
\eeq
of functionals $A[\phi,\pi]$ of the phase
space variables $\phi^i(x)$ and
$
\pi_i(x)=\frac{\partial \cal L}{\partial \partial_0\phi^i}(x)
$
by the graded Poisson bracket
\bea
\nonumber
&\{A,B\}_{Poisson}= &\\
&\int d^3x
\left( (-1)^{|i|(|i|+|A|)}
\frac{\delta A}{\delta \phi^i(x)}\frac{\delta B}{\delta \pi_i(x)}
- (-1)^{|i||A|}
\frac{\delta A}{\delta \pi_i(x)}\frac{\delta B}{\delta \phi^i(x)}
\right)\ .
\eea
If one investigates the quantized theory then in the simplest of all
conceivable worlds the classical Poisson brackets
would be replaced by (anti-) commutators of quantized operators.
In particular the BRS  operator $Q_s$ would commute with the scattering
matrix $S$
\beq
\label{s-matrix}
S=``\,{\rm T}\,e^{i\int d^4x{\cal L}_{int}}\," \qquad [Q_s,S]=0
\eeq
and scattering processes would map physical states unitarily to physical
states
\beq
S {\cal H}_{phys} = {\cal H}_{phys}\ .
\eeq
Classically an invariant action is sufficient to ensure this property.
The perturbative evaluation of scattering amplitudes, however, has to
face the problem that the $S$-matrix (\ref{s-matrix}) has ill defined
contributions from products of ${\cal L}_{int}(x_1)\dots{\cal L}_{int}(x_n)$
if arguments $x_i$ and $x_j$ coincide. Though upon integration
$\int dx_1\dots dx_n$ this is a set of measure zero
these products of fields at coinciding space time arguments are the 
reason for all divergencies which emerge upon the
naive application of the Feynman rules. More precisely the $S$-matrix is
a time ordered series in $i\int d^4x{\cal L}_{int}$ and a set of
prescriptions (indicated by the quotes in (\ref{s-matrix})) to define
in each order the products of ${\cal L}_{int}(x)$ at coinciding space-time
points. To analyze these divergencies it is sufficient to consider only
connected diagrams. In momentum space they 
decompose into products of one particle irreducible
$n$-point functions $\tilde{G}_{1PI}(p_1,\dots,p_n)$ which define the 
effective action \bea
\Gamma[\phi]& = &\sum_{n=0}^{\infty}\frac{1}{n!}\int d^4x_1\dots d^4x_n\,
\phi(x_1)\dots\phi(x_n)\,G_{1PI}(x_1,\dots,x_n)\\
& = & \int d^4x {\cal L}(\phi,\partial \phi, \dots ) +
\sum_{n\geq 1}\hbar^n\Gamma_n[\phi]
\eea
To lowest order in $\hbar$ the effective action $\Gamma$ is given by the
classical action $\Gamma_0 = \int d^4x {\cal L}_0$. This is a local functional,
in particular ${\cal L}_0$ is a series in the fields and a polynomial in the
partial derivatives of the fields. The Feynman diagrams fix the expansion
of the nonlocal effective action $\Gamma = \sum \hbar^n \Gamma_n$ 
up to local functionals which can be chosen
in each loop order. We are free to choose the Lagrange density in each 
loop order, i.e as a series in $\hbar$.
\beq
{\cal L}= {\cal L}_0 + \sum_{n\geq 1}\hbar^n{\cal L}_n
\eeq
Consider in each loop order the question whether the full effective action is
BRS  invariant.
$$
s\, \Gamma[\phi] =0
$$
To lowest order in $\hbar$ this requires the Lagrange density 
${\cal L}_0$ to be a solution of (\ref{relcoh}). Assume that one has
satisfied $s\, \Gamma[\phi] =0$ up to $n$-loop order.

The naive calculation of
$n+1$-loop diagrams contains divergencies which make it necessary to
introduce a regularization, e.g. the Pauli-Villars regularization,
and counterterms  (or use a prescription such as dimensional regularization
or the BPHZ prescription which is a shortcut for regularization and
counterterms).
No regularization respects locality, unitarity and symmetries simultaneously,
otherwise it would not be a regularization but an acceptable theory.
The Pauli-Villars regularization is local. It violates unitarity for energies
above the regulator masses and also because it violates BRS  invariance. 
If one cancels the
divergencies of diagrams with counterterms and considers the limit of infinite
regulator masses then unitarity is obtained if the BRS  symmetry guarantees
the decoupling of the unphysical gauge modes. Locality was preserved for
all values of the regulator masses. What about BRS  symmetry?

One cannot argue
that one has switched off the regularization and that therefore the symmetry
should be restored.  There is the phenomenon of hysteresis.
For example: if you
have a spherically symmetric iron ball and switch on a symmetry breaking
magnetic field then the magnetic properties of the iron ball will
usually not become spherically symmetric again if the magnetic field is
switched off.
Analogously in the calculation of 
$\Gamma_{n+1}$ we have to be prepared that the regularization and the 
cancellation of divergencies
by counterterms does not lead to an invariant effective action but rather to
\beq
s\,\Gamma = \hbar^{n+1}a + \sum_{k\geq n+2}\hbar^k a_k\ .
\eeq

If the  functional $a$ cannot be made to vanish by an appropriate
choice of ${\cal L}_{n+1}$ then the BRS  symmetry is broken by the
anomaly $a$. 

Because $s$ is nilpotent the anomaly $a$ has to satisfy
\beq
\label{consis}
s\,a=0\ .
\eeq

This is the celebrated consistency condition of Wess and Zumino 
\cite{wesszumino}.
The consistency condition has acquired an outstanding importance because it
allows to calculate all possible anomalies $a$ as the general solution to
$s\,a =0$ and to check in each given model whether the anomaly actually
occurs. At first sight one would not expect that the consistency equation
has comparatively few solutions. The BRS  transformation of arbitrary 
functionals
satisfies $s\,a=0$. The anomaly $a$, however, arises from the divergencies
of Feynman diagrams where all subdiagrams are finite and compatible with
BRS  invariance. These divergencies can be isolated in parts of the
$n$-point functions which depend polynomially on the external momenta, 
i.e. in local functionals. Therefore it turns out that the anomaly is a
local functional.
\beq
a = \int d^4x {\cal A}^1(\phi(x),\partial \phi(x),\dots)
\eeq
The anomaly density ${\cal A}^1$ is a series in the
fields $\phi$ and a polynomial in the partial derivatives 
of the fields comparable to a Lagrange density but with ghost number +1.
The integrand ${\cal A}^1$ represents an equivalence class. It is
determined only up to terms of the form
$s{\cal L}$ because
we are free to choose contributions to the Lagrange density at each loop
order, in particular we try to choose ${\cal L}_{n+1}$ such that
$s{\cal L}_{n+1}$ cancels ${\cal A}^1$ in order to make
$\Gamma_{n+1}$ BRS  invariant. Moreover ${\cal A}^1$ is determined
only up to derivative terms of the form $d\eta^1$.

${\cal A}^1$ transforms into a derivative because the
anomaly $a$ satisfies the consistency condition. We combine the
anomaly density ${\cal A}^1$ with $d^Dx$ to a volume form
$\omega^1_{D}$ and denote the ghost numbers as superscripts and the 
form degree as subscript. Then the
consistency condition  and the description of the equivalence class read
\beq
\label{anomaly}
s \omega^1_{D} + d\omega^2_{D-1} = 0 \quad \omega^1_{D} 
\,\bmod\,s\eta^0_{D} + d \eta^1_{D-1}\ . 
\eeq
This equation determines all possible anomalies and can be 
analyzed if one is given the field content and the 
BRS  transformations $s$. Its solutions do not depend on particular 
properties of the model under consideration.

The determination of all possible
anomalies is again a cohomological problem just as the determination
of all BRS  invariant local actions (\ref{relcoh}) but now with ghost numbers
shifted by +1.
We will deal with both equations and consider the equation
\beq
\label{rel}
s \omega^g_{D} + d\omega^{g+1}_{D-1} = 0 \quad \omega^g_{D}
\,\bmod\,s\eta^{g-1}_{D} + d \eta^g_{D-1} \eeq
for arbitrary ghost number $g$. The form degree is given by the 
subscript.


\chapter{Cohomological Problems}
In the preceding chapters we have encountered repeatedly
the cohomological problem to solve the linear equation
$s\omega = 0\quad \omega \bmod s\eta$
where $s$ is a nilpotent operator $s^2=0$. The equivalence classes of
solutions $\omega$ form a linear space, the cohomology $H(s)$ of $s$.
The equivalence classes of solutions $\omega^g_p$ of the problem 
$s\omega^g_p = - d\omega^{g+1}_{p-1} \quad \omega^g_p \bmod 
d\eta^g_{p-1} + s\eta^{g-1}_{p}$, where $s^2=0=d^2=\{s,d\}$ form
the relative cohomology $H^g_p(s|d)$ of $s$ modulo $d$ of ghost number 
$g$ and form degree $p$.

Let us start to solve such equations and consider the problem to determine
the physical multiparticle states. Multiparticle states can be written as
a polynomial $P$ of the creation operators acting  on the vacuum
$$
P(a^\dagger_\tau,c^\dagger,\bar{c}^\dagger)|0\rangle
\quad \tau =k,\bar{k},1,2
$$
if one neglects the technical complication that all these creation operators
depend on $\vec{k}$ and have to be smeared with normalizable functions.
The BRS operator $Q_s$ acts on these states in the same way as the 
algebraic operation 
\beq
s=\sqrt{2}|\vec{k}|
(ia^\dagger_{\bar{k}}\frac{\partial}{\partial\bar{c}^\dagger} +
c^\dagger\frac{\partial}{\partial a^\dagger_k})
\eeq
acts on polynomials in commuting and anticommuting variables. For
one particle states, i.e. linear homogeneous polynomials $P$ we had 
concluded
that the physical states, the cohomology of $Q_s$ with particle number 
1, are generated by the transverse creation operators
$a^\dagger_i$, i.e. by variables which are neither generated by $s$ such
as  $a^\dagger_{\bar{k}}$ 
or $c^\dagger$ nor transformed such as $\bar{c}^\dagger$ and 
$a^\dagger_k$. Let us systematize our notation and
denote the variables collectively by $x^m$. Then the operator $s$
becomes the nilpotent operator $d$ ((\ref{dext}) without reality 
property). It maps the variables $x^m$ to $dx^m$ with opposite
statistics.
\beq
d=dx^m\frac{\partial\phantom{x^m}}{\partial x^m}
\quad |dx^m|= |x^m| +1
\eeq
We claim that on polynomials in $x^m$ and $dx^m$ the cohomology of the
exterior derivative $d$ is described by the basic lemma.
\begin{theorem}{Basic Lemma}
\beq
\label{basic}
d f(x,dx)=0 \Leftrightarrow f(x,dx)= f_0 + d g(x,dx)\ .
\eeq
\end{theorem}
$f_0$ denotes the polynomial which is homogeneous of degree 0 in
$x^m$ and $dx^m$ and is therefore independent of these variables.

Applied to the  Fock space the Basic Lemma implies that physical 
$n$-particle states are generated by polynomials $f_0$ of creation 
operators which contain no operators 
$a^\dagger_{\bar{k}},a^\dagger_k,c^\dagger,\bar{c}^\dagger $.
Physical states are generated from the transverse creation operators
$a^\dagger_i,\ i=1,2$.

To prove the lemma we introduce the operation
\beq
\delta = x^m\frac{\partial}{\partial (dx^m)}\ .
\eeq
The anticommutator $\Delta$ of $d$ and $\delta$ counts the 
variables $x^m$ and $dx^m$.
\beq
\label{delta}
\{d,\delta\}=\Delta = x^m\frac{\partial}{\partial x^m} +
dx^m\frac{\partial}{\partial (dx^m)}= N_x + N_{dx}
\eeq
From the relation $d^2=0$ it follows that $d$ commutes with 
$\{d,\delta\}$.
\beq
\label{danti}
d^2=0 \Rightarrow [d,\{d,\delta\}]=0
\eeq
Of course we can easily check explicitly that $d$ does not change the 
number 
of variables $x$ and $dx$ in a polynomial. We can decompose each polynomial
$f$ into pieces $f_n$ of definite homogeneity $n$ in the variables $x$ and
$dx$, i.e. $(N_x+N_{dx})f_n = n f_n$. Using (\ref{delta}) we can write 
$f$ in the following form. \bea
\nonumber
f&=&f_0 + \sum_{n\ge 1} f_n =
 f_0 + \sum_{n\ge 1} (N_x+N_{dx})\frac{1}{n}f_n\\
\nonumber
&=&f_0 + d \left( \delta \sum_{n\ge 1} \frac{1}{n} f_n \right)
+ \delta \left( d \sum_{n\ge 1} \frac{1}{n} f_n \right) \\
\label{hodge}
f &=& f_0 +d\eta + 
\delta \chi \eea
This is the Hodge decomposition of an arbitrary polynomial in $x$ and $dx$ into
a zero mode $f_0$, a $d$-exact part $d\eta$ and a $\delta$-exact part 
$\delta\chi$. If $f$ solves $df=0$
then the equations $d f_n = 0$ have to hold for each piece $d f_n$
separately because $d$ commutes
with the number operator $\Delta$. But $d f_n =0 $ implies that the last term
in the Hodge decomposition, the $\delta$-exact term, vanishes. This 
proves our
lemma. Of course this is not our lemma: it is  Poincar\'e\,'s lemma for forms
in a star shaped domain if one writes $\frac{1}{n}$ as
$\int_0^1 dt\, t^{n-1}$.
\beq
\nonumber
f(x,dx)=f_0 + \{d,\delta\}\int_0^1\frac{dt}{t}(f(tx,tdx)-f(0,0))
\eeq
\begin{theorem}{Poincar\'e\,'s lemma}
\label{poinc}
\beq
df(x,dx)=0 \Leftrightarrow f(x,dx)=f_0 + d 
\,\delta\int_0^1\frac{dt}{t}(f(tx,tdx)-f(0,0))
\eeq
\end{theorem}
In this form the lemma is not restricted to polynomials but applies to 
arbitrary differential forms
$f$ which are defined along the ray $tx$ for 
$0\le t \le 1$. Note that the integral is not 
singular at $t=0$ .
 
We chose to present the Poincar\'e lemma in the algebraic form --
though it applies only to polynomials and to analytical functions if 
one neglects the question of convergence -- because
we will follow a related strategy to solve the cohomological problems to come:
given a nilpotent operator $d$ we inspect operators $\delta$ and their
anticommutators $\Delta$ and try to invert $\Delta$. Only the zero modes
of $\Delta$ can contribute to the cohomology of $d$.

We have to generalize Poincar\'e$\,$'s lemma because we
consider Lagrange densities
and more generally forms $\omega$ which are series in fields $\phi$, 
polynomials in derivatives of fields
$\partial_m \phi,\,\dots, \partial_{m_1}\dots \partial_{m_l}\phi$,
polynomials in $dx^m$ and series in the coordinates $x^m$.
\beq
\omega = \omega(x,dx,\phi,\partial \phi, \partial \partial \phi,\dots )
\eeq
Such forms occur as integrands of local functionals. Because they depend
polynomially on derivatives of fields they contain only terms with a
bounded number of derivatives, though there is no bound on the
number of derivatives which is common to all forms $\omega$.
We call the fields and their derivatives
\beq 
\{\phi\}=\phi,\partial \phi, \partial \partial \phi,\dots
\eeq
the jet variables. Poincar\'e$\,$'s lemma does not apply to forms which 
depend on the coordinates, the differentials and the jet variables. The 
exceptions are
Lagrange densities which lead to nontrivial Euler-Lagrange equations.
Then the Lagrange density $\omega = {\cal L}d^D x$ 
cannot be a total derivative $\omega \neq d\eta$ though $d\omega = 0$
because $\omega$ is a volume form. 
Let us prove this result.

The exterior derivative on forms of jet variables differentiates the 
explicit coordinates $x^m$, the fields just get an additional label upon
differentiation
\beq
d = dx^m \partial_m \quad
\partial_m x^n = \delta_m^{\phantom{m}n}
\quad \partial_k (\partial_l\dots \partial_m \phi) = 
\partial_k \partial_l\dots \partial_m \phi\ .
\eeq
The fields are assumed to satisfy no differential equation, i.e. the variables
$\partial_k \partial_l\dots \partial_m \phi$ are independent up to the fact that 
partial derivatives commute
$\partial_k \dots \partial_m \phi$=
$\partial_m \dots \partial_k \phi\ .$
On such variables we can define the operation $t^n$
\beq
t^n (\partial_{m_1}\dots\partial_{m_l}\phi)=
\sum\limits_{i=1}^l \delta_{m_i}^n \partial_{m_1}\dots\hat{\partial}_{m_i}\dots
\partial_{m_l}\phi
\quad t^n(x^m)=0 \quad t^n(dx^m)=0
\eeq
The hat $\hat{}$ means omission of the hatted symbol. We define the
action of $t^n$ on polynomials in the jet variables by linearity and
the Leibniz rule. $t^n$ acts on derivatives of the fields $\phi$ like a 
differentiation with respect
to $\partial_n$, i.e. $t^n=\frac{\partial}{\partial(\partial_n)}$.
Obviously one gets $[t^m,t^n] =0$ from this definition. Less trivial is
\beq
\label{td}
[t^n,\partial_m]=\delta_m^n N_{ \{\phi\} }
\eeq
$N_{ \{\phi\}}$ counts the jet variables $\{\phi\}$. The
equation holds for linear polynomials, i.e. for the jet variables and 
coordinates and differentials, and extends to arbitrary polynomials 
because both sides of this equation satisfy the Leibniz rule.

To determine the cohomology of $d=dx^m\partial_m$ we consider 
separately forms $\omega$ with a fixed form degree $p$:
\beq
N_{dx}=dx^m\frac{\partial}{\partial(dx^m)}\quad N_{dx}\omega
= p \,\omega\ .
\eeq
which are homogeneous of degree $N$ in  $\{\phi\}$. We assume $N > 0$, 
the case $N=0$ is covered by Poincar\'e$\,$'s lemma 
(theorem \ref{poinc}).

Consider the operation 
\beq
b=t^m\frac{\partial}{\partial(dx^m)}
\eeq
and calculate its anticommutator with the exterior derivative $d$
as an exercise in graded commutators:
\bea
\nonumber
\{b,d\}&=&\{t^m\frac{\partial}{\partial (dx^m)},dx^n\}\partial_n
-dx^n[t^m\frac{\partial}{\partial (dx^m)},\partial_n]\\
\nonumber
&=&t^m\delta_m^{\phantom{m}n}\partial_n - dx^n\delta_n^{\phantom{n}m}N
\frac{\partial}{\partial dx^m}\\
\nonumber
&=& \partial_nt^n + \delta_n^{\phantom{n}n}N - N\,N_{dx}\ .\\
\eea
So we get
\beq
\label{db}
\{d,b\}=N(D-N_{dx}) + P_1\ .
\eeq
$D=\delta_n^n$ is the dimension of the manifold, the operator $P_1$ is
given by
\beq
P_1=\partial_kt^k\ .
\eeq

Consider more generally the operations $P_n$
\beq
P_n=\partial_{k_1}\dots\partial_{k_n}t^{k_1}\dots t^{k_n}
\eeq
which take away $n$ derivatives and redistribute them afterwards.
For each polynomial $\omega$ in the jet variables there exists a 
$\bar{n}(\omega)$
such that 
\beq
\label{maxn}
P_n \omega =0 \;\forall n \ge \bar{n}(\omega)
\eeq
 because each monomial of
$\omega$ has a bounded number of derivatives.
Using the commutation relation (\ref{td}) one proves the recursion 
relation
\beq
P_1P_k=P_{k+1}+kNP_k
\eeq
which can be used iteratively to express $P_k$ in terms of $P_1$ and $N$
\beq
\label{pk} 
P_k=\prod\limits_{l=0}^{k-1}(P_1-lN)\ .
\eeq

Using the argument (\ref{danti}) that a nilpotent operation commutes 
with all its anticommutators we conclude  from (\ref{db})
\beq
[d,N(D-N_{dx}) + P_1] =0\ .
\eeq

Therefore $d\omega = 0$ implies $d(P_1\omega) = 0$ and from (\ref{pk})
we conclude $d(P_k\omega) =0$ by induction. We use the relation 
(\ref{db})
to express these closed forms $P_k\omega$ as exact forms up to terms
$P_{k+1}\omega$.
\bea
\nonumber
d(b\omega)   &=& P_1\omega + N(D-p)\omega\\
\nonumber
 d(bP_k\omega) &=&P_1P_k\omega + N(D-p)P_k\omega\\
\nonumber
               &=&P_{k+1}\omega +kNP_k\omega + N(D-p)P_k\omega\\
\label{pkomega}
d(bP_k\omega) &=& P_{k+1}\omega + N(D-p+k)P_k\omega \quad k=0,1,\dots
\eea

If $p<D$ then we can solve for $\omega$ in terms of exact forms 
$d(b\omega)$ and $P_{1}\omega$ which can be expressed as exact form and 
a term $P_2\omega$ and so on. This recursion terminates because 
$P_n\omega =0 \;\forall n \ge \bar{n}(\omega)$ (\ref{maxn}).
Explicitly we have for $p<D$ and $N>0$:
\beq
\label{pkd}
d\omega = 0 \Rightarrow \quad
\omega = d\left ( b 
\sum\limits_{k=0}^{\bar{n}(\omega)}\frac{(-)^k}{N^{k+1}}
\frac{(D-p-1)!}{(D-p+k)!}P_k\omega \right ) = d \eta\ .
\eeq

To complete the investigation of the cohomology of $d$ we have to 
consider volume forms $\omega = {\cal L}d^Dx$. We treat separately
pieces ${\cal L}_N$ which are homogeneous of degree $N>0$ in the jet 
variables $\{\phi\}$. These pieces satisfy
\bea
\label{vari}
\nonumber
N{\cal L}_N&=&\phi^i
\frac{\partial{\cal L}_N}{\partial \phi^i} +
\partial_m\phi^i\frac{\partial{\cal L}_N}{\partial (\partial_m\phi^i)}
+ \dots\\
&=& \phi^i \frac{\hat{\partial}{\cal L}_N}{\hat{\partial} \phi^i} + 
\partial_m X_N^m 
\quad 
X_N^m =  \phi^i\frac{\partial{\cal L}_N}{\partial (\partial_m\phi^i)} + 
\dots\ .
\eea
Here we use the notation
\beq
\label{eulder}
\frac{\hat{\partial}{\cal L}}{\hat{\partial} \phi^i}
= \frac{\partial{\cal L}}{\partial \phi^i} 
-\partial_m \frac{\partial{\cal L}}{\partial (\partial_m\phi^i)}
+ \dots
\eeq
for the Euler derivative of the Lagrange density. The dots denote terms 
which come from higher derivatives. The derivation of (\ref{vari}) 
is analogous to the derivation of the Euler Lagrange equations from the
action principle. For the volume form $\omega_N ={\cal L}_Nd^Dx$ 
 (\ref{vari}) reads 
\beq
{\cal L}_Nd^Dx = \frac{1}{N}\phi^i 
\frac{\hat{\partial}{\cal L}_N}{\hat{\partial} \phi^i} d^Dx +
d \left ( \frac{1}{N}X_N^m\frac{\partial}{\partial (dx^m)}d^Dx\right )
\eeq
If we combine this equation with Poincar\'e$\,$'s lemma 
(theorem \ref{poinc})
and with (\ref{pkd}), combine terms with different degrees of 
homogeneity $N$ and different form degree $p$ we obtain the Algebraic 
Poincar\'e Lemma for forms of the coordinates, differentials and jet
variables
\begin{theorem}{Algebraic Poincar\'e Lemma}
\label{algebraic}
\beq
d \omega(x,dx,\{\phi\}) = 0 \Leftrightarrow \omega(x,dx,\{\phi\})
 = const + d\eta(x,dx,\{\phi\}) + {\cal L}(x,\{\phi\})d^Dx 
\eeq
\end{theorem}
The Lagrange form ${\cal L}(x,\{\phi\})d^Dx$ is trivial, i.e. of the 
form $d\eta$, 
if and only if its Euler derivatives with respect to all fields vanish.

The Algebraic Poincar\'e Lemma does not hold if the base manifold is not
starshaped or if the fields $\phi$ take values in a 
topologically nontrivial target space. In these cases the 
operations $\delta = x\frac{\partial}{\partial (dx)}$ and
$b=t^n\frac{\partial}{\partial (dx^n)}$ cannot be 
defined because
a relation like $x\cong x+2\pi$, which holds for the coordinates on a 
circle, would lead to the contradiction 
$0\cong 2\pi \frac{\partial}{\partial (dx)}$. Here we restrict our 
investigations to topologically trivial base manifolds and topologically 
trivial target spaces. It is the 
topology of the invariance groups and the Lagrangean solutions in 
the Algebraic Poincar\'e lemma which give rise to a nontrivial cohomology of 
the exterior derivative $d$ and the BRS transformation $s$.

The Algebraic Poincar\'e lemma is modified if the jet space 
contains in addition variables which are space time constants. This 
occurs for example if one treats rigid transformations as 
BRS transformations with constant ghosts $C$, i.e. $\partial_m C =0$.
If these ghosts occur as variables in forms $\omega$ then they are not
counted by the number operators $N$ which have been used in the proof of 
the Algebraic Poincar\'e Lemma and can appear as variables in
$const=f(C)$, in $\eta$ and in ${\cal L}$.

We are now prepared to investigate the relative cohomology
and derive the so called descent equations. We recall that we deal with
two nilpotent derivatives, the exterior derivative $d$ and the 
BRS transformation $s$, which anticommute which each other
\beq
\label{anticom}
d^2 =0 \quad s^2=0 \quad \{s,d\}=0\ .
\eeq
$s$ leaves the form degree $N_{dx}$ invariant, $d$ raises it by 1
\beq
\label{Ndegree}
[N_{dx},s] =0 \quad [N_{dx},d]=d\ .
\eeq

We consider the equation
\beq                                                   
\label{relcoh1}                                         
s\,\omega_D + d\,\omega_{D-1} = 0\quad                     
\omega_D \ \bmod\ ( s\eta_D + d\eta_{D-1} )\ .          
\eeq                                                   
The subscript denotes the form degree. The relative cohomology 
(\ref{relcoh1}) relates forms of different ghost number
\beq
gh(\omega_{D})=gh(\omega_{D-1})-1=gh(\eta_D)+1=gh(\eta_{D-1})
\eeq
Let us derive the descent equations as a necessary consequence of 
(\ref{relcoh1}). We apply $s$ and use (\ref{anticom})
\beq
0=s(s\,\omega_D + d\,\omega_{D-1})=sd\,\omega_{D-1}=d(- s\,\omega_{D-1})
\ .\eeq
By the Algebraic Poincar\'e Lemma (\ref{algebraic}) $- s\,\omega_{D-1}$ 
is of the form $const + d\eta(\{\phi\}) + {\cal L}(\{\phi\})d^Dx$. 
The piece ${\cal L}(\{\phi\})d^Dx$ has to vanish because $\omega_{D-1}$
has form degree $D-1$ and if $D>1$ then also the piece $const$ vanishes.
Therefore we conclude
\beq                                                   
\label{relcoh2}                                        
s\,\omega_{D-1} + d\,\omega_{D-2} = 0\quad                 
\omega_{D-1} \ \bmod\ ( s\eta_{D-1} + d\eta_{D-2} )        
\eeq                                                   
where we denoted $\eta$ by $\omega_{D-2}$ to indicate its form degree. 
Adding to $\omega_{D-1}$ a piece of the form $s\eta_{D-1}+d\eta_{D-2}$
changes $\omega_D$ only within its class of equivalent representatives. 
Therefore $\omega_{D-1}$ is naturally a representative of an equivalence
class.
From (\ref{relcoh1}) we have derived (\ref{relcoh2}) which is nothing 
but (\ref{relcoh1}) with  form degree  lowered by 1.
Iterating the arguments we lower the form degree step by step and obtain 
the descent equations
\beq                                                           
\label{absteig}                                                
s\,\omega_{i} + d\,\omega_{i-1} = 0\quad i=D,D-1,\dots,1\quad                    
\omega_{i} \ \bmod\ ( s\eta_{i} + d\eta_{i-1} )\
\eeq                                                           
until the form degree drops to zero. It cannot become 
negative. For $i=0$ one has
\beq
\label{scoh}
s\,\omega_0 = 0 \quad 
\omega_0 \ \bmod\ s\eta_0\ .
\eeq
A more careful application of the Algebraic Poincar\'e Lemma 
would only have allowed to conclude  
$$s\ \omega_0 = const\ .$$
If, however, the BRS transformation is not spontaneously broken
i.e. if $s \phi  {}_{|(\phi=0)}=0$ then $s\ \omega_0$ has to vanish.
This follows most easily if one evaluates  both sides of $s\ \omega_0 = 
const$ for vanishing fields. We assume for the following that the 
BRS transformations are not spontaneously broken.
We will exclude from  our considerations also
spontaneously broken rigid symmetries. There we cannot apply these 
arguments because then $s \phi  {}_{|(\phi=0)}=C$ gives ghosts which are
space time constant and one can have 
$s\ \omega_0 = const = f(C) \neq  0$.

Actually the descent equations (\ref{absteig},\ref{scoh}) are just 
another cohomological equation for a nilpotent operator
$\tilde{s}$ and
a form $\tilde{\omega}$
\beq
\label{stilde}
\tilde{s} = d + s \quad \tilde{s}^2=0
\eeq
\beq
\label{otilde}
\tilde{\omega }= \sum\limits_{i=0}^{D}\omega_i 
\eeq
\beq
\label{stildecoh}
\tilde{s}\ \tilde{\omega } = 0\quad 
\tilde{\omega } \bmod \tilde{s} \tilde{\eta}\ .
\eeq
The fact that $\tilde{s}$ is nilpotent follows from (\ref{anticom}). The
descent equations (\ref{absteig}, \ref{scoh})  imply 
$\tilde{s}\ \tilde{\omega } = 0$. The equivalence class of 
$\tilde{\omega}$ is given by $\tilde{s}(\sum_i \eta_i)$. So 
(\ref{stildecoh}) is a consequence of the descent equations. On the 
other hand if (\ref{stilde}) holds then the equation (\ref{stildecoh}) 
implies the descent equations. This follows if one splits
$\tilde{s}$, $\tilde{\omega}$ and  $\tilde{\eta}$ with respect to the 
form degree (\ref{Ndegree}).

Let us formulate this result as
\begin{theorem}{\quad}\\
\label{absts}
If $\tilde{s}=s+d$ is a sum of two fermionic operators
where $s$ preserves
the form degree and $d$ raises it by one, then $\tilde{s}$ is nilpotent
if and only if $s$ and $d$ are nilpotent and anticommute. 

Each solution $ (\omega_0,\dots, \omega_D )$, $\omega_i 
\bmod s\ \eta_i + d\ \eta_{i-1}$ of the descent equations 
(\ref{absteig},
\ref{scoh}) with nilpotent, anticommuting operators $s$ and $d$
corresponds one to one to an element $\tilde{\omega}$ of the cohomology 
$H(\tilde{s})= \{\tilde{\omega}: \ \tilde{s}\ \tilde{\omega } = 0\quad   
\tilde{\omega } \bmod \tilde{s} \tilde{\eta})\}$.
$\omega_i$ are the parts of $\tilde{\omega}$ with form degree $i$.
\end{theorem}

The formulation of the descent equations as a cohomological problem of
the operator $\tilde{s}$ has several virtues. The solutions to 
$\tilde{s}\ \tilde{\omega } = 0$ 
can obviously be multiplied to obtain further solutions. Phrased 
mathematically they form an algebra not just a vector space.
More importantly for the BRS operator in gravitational Yang Mills 
theories we will find that the equation $\tilde{s}\ \tilde{\omega } = 0$
can be cast into the form $s\ \omega = 0$ by a change of variables,
where $s$ is the original BRS operator. This
equation has to be solved anyhow as part of the descent equations. Once 
one has solved it one can recover the complete solution of the descent 
equations, in particular one can read off $\omega_D$ as the $D$ form 
part of $\tilde{\omega}$. These virtues
justify to consider with $\tilde{\omega}$ a sum of forms of different 
form degrees which in traditional eyes would be considered to add 
peaches and apples.

As the last subject of this chapter we study the action of a
nilpotent derivative $d$ on a product $A=A_1\times A_2$ of 
vectorspaces (algebras) which are separately invariant under $d$
\beq
d A_1 \subset A_1 \quad d A_2 \subset A_2 \ .
\eeq
Knneth's theorem states  that the cohomology $H(A,d)$ of $d$ 
acting on $A$ is given by the product of the cohomology
$H(A_1,d)$ of $d$ acting on $A_1$
and $H(A_2,d)$ of $d$ acting on $A_2$. 

\begin{theorem}{K\"unneth-formula}\\
\label{kunn}
Let $d=d_1+d_2$ be a sum of nilpotent differential 
operators which leave their vectorspaces $A_1$ and $A_2$ invariant
\beq
d_1 A_1 \subset A_1\quad  d A_2 \subset A_2
\eeq
and which are defined on the product $A=A_1\times A_2$ by the Leibniz 
rule \beq
d_1 (kl) = (d_1 k)l \quad d_2(kl) = (-)^{|k|}k (d_2 l)\quad
\forall k\in A_1,\: l\in A_2\ .
\eeq
Then the  cohomology $H(A,d)$ of $d$ acting on $A$ is the 
product of the cohomologies of $d_1$ acting on $A_1$ and $d_2$ acting
on $A_2$
\beq
H (A_1\times A_2,d_1+d_2)=H (A_1,d_1)\times H (A_2,d_2)
\eeq
\end{theorem}

To prove the theorem we consider an element $f\in H(d)$
\beq
f=\sum_ik_il_i
\eeq
given as a sum of products of elements $k_i\in A_1$ and $l_i\in A_2$.
Without loss of generality we assume that the elements $k_i$
are taken from a basis of $A_1$ and the elements $l_i$ 
are taken from a basis of $A_2$.

\bea
\label{link}
\sum c_i k_i &=& 0 \Leftrightarrow c_i = 0  \ \forall\  i\\
\label{linl}
\sum c_i l_i &=& 0 \Leftrightarrow c_i = 0 \ \forall\  i
\eea
Otherwise one has a relation like
$l_1=\sum^\prime_i \alpha_i l_i$ or 
$k_1=\sum^\prime_i \beta_i k_i$,
where $\sum^\prime$ does not contain $i=1$, and can rewrite $f$
with fewer terms 
$f=\sum^\prime_i(k_i+\alpha_ik_1) \cdot l_i $ or
$f=\sum^\prime_i k_i\cdot
(l_i + \beta_il_1)$.
We can even choose $f\in H(d)$  in such a manner that the elements $k_i$ 
are taken from a basis of a complement to the space $d_1 A_1$. In other
words we can choose $f$ such that no linear combination of the elements
$k_i$ combines to a $d_1$-exact form.
\beq
\sum_i c_i k_i = d_1g \Leftrightarrow d_1g = 0 =c_i  \ \forall\  i
\eeq
Otherwise we have a relation like
$k_1=\sum^\prime_i \beta_i k_i +  d_1 \kappa$,
where $\sum^\prime$ does not contain $i=1$, and we can rewrite 
$f\in H(d)$ up to an irrelevant piece $d(\kappa\ l_1) $
$f=\sum^\prime_i k_i\cdot (l_i + \beta_il_1) - (-)^{|\kappa|}  \kappa \ 
d_2 l_1 + d(\kappa\  l_1)$ with elements $k^\prime_i=\kappa,k_2,\dots$ .
We can  iterate this argument until no linear combination of the 
elements $k^\prime_i$ combines to a $d_1$-exact form.

By assumption $f$ solves $df\,=\,0$ which implies
\beq
\sum_i \left ((d_1k_i)l_i + (-)^{k_i}k_i (d_2 l_i) \right ) = 0 \ .
\eeq
In this sum $\sum_i (d_1k_i)l_i$ and $\sum_i (-)^{k_i}k_i (d_2 l_i)$ 
have to vanish separately because the elements $k_i$ are linearly
independent from the elements $d_1 k_i\in \, d_1A_1$. 
$\sum_i (d_1k_i)l_i =0 $, however implies 
\beq d_1k_i = 0
\eeq
because the 
elements $l_i$ are linearly independent and
$\sum_i (-)^{k_i}k_i (d_2 l_i)=0$ leads to
\beq
d_2l_i = 0
\eeq
analogously. So we have shown
\beq
df=0 \Rightarrow f=\sum_i k_i l_i + d\chi \mbox{ where }
d_1 k_i = 0 = d_2 l_i \ \forall i \ .
\eeq
Changing $k_i$ and $l_i$ within their equivalence class 
$k_i \bmod d_1\kappa_i$ and $l_i \bmod d_2\lambda_i$ does not change the
equivalence class $f \bmod d\chi$:
\beq
\sum_i(k_i + d_1\kappa_i)(l_i + d_2\lambda_i) = \sum_i k_i l_i +
d\ \sum_i \left (\kappa_i (l_i + d_2 \lambda_i) + (-)^{k_i}k_i \lambda_i
\right )
\eeq
Therefore $H(A, d)$ is contained in $H_1 (A_1, d_1)\times H_2 (A_2, 
d_2)$.
The inclusion $H_1 (A_1, d_1)\times H_2 (A_2, d_2) \subset H(A,d)$ is 
trivial. This concludes the proof of the theorem.


\chapter{BRS algebra of Gravitational Yang Mills Theories}
Gauge theories such as gravitational Yang Mills theories theories rely 
on tensor analysis.  The set of tensors is a subalgebra of 
the polynomials in the jet variables.
\beq
\left ( Tensors\right )
\subset 
\left ( Polynomials(\phi,\partial\phi, \partial \partial \phi,\dots 
)\right ) \eeq
The covariant operations $\Delta_M$ which are used in tensor analysis
\beq
\Delta_M: \left (Tensors\right ) \rightarrow \left (Tensors\right )
\eeq
map tensors to tensors and satisfy the Leibniz rule (\ref{leibniz}).
These covariant operations have a basis consisting of the covariant 
space time derivatives $D_a,\ a=0,\dots,D-1$ and 
spin and isospin transformations $\delta_I$, which correspond to a basis 
of
the Lie algebra of the Lorentz group and of the gauge group, and - if 
one considers supergravitational theories - the covariant spinor 
derivatives $D_\alpha$, $D_{\dot{\alpha}}$ .
\beq
( \Delta_M )= (D_a,\ \delta_I,\ D_\alpha, D_{\dot{\alpha}} )
\eeq
The space of covariant operations is closed with respect to 
graded commutation 
\beq
\label{diffa}
[\Delta_M,\Delta_N] := \Delta_M\Delta_N - (-)^{MN}\Delta_N\Delta_M
={\cal F}_{MN}^{\phantom{MN}K}\Delta_K \ .
\eeq
The structure functions $\F{MN}{K}$ are also tensors. Some of 
these structure functions have purely numerical values as for example 
the structure constants of the spin and isospin Lie algebra
\beq
\label{diffiso}
[\delta_I,\delta_J] = \f{IJ}{K}\delta_K
\eeq
or the matrix elements of representations of the Lorentz algebra
\beq
[\delta_{[a,b]},D_c]= -(G_{[a,b]})_c^{\phantom{c}d}D_d
=\eta_{ca}D_b - \eta_{cb}D_a
\eeq
or constant torsion in superspace.
Other components of the tensors $\F{MN}{K}$ are given by the Riemann 
curvature, the Yang-Mills field strength and in supergravity the 
Rarita-Schwinger field strength and auxiliary fields of the 
supergravitational multiplet. We use the word field strength also to 
denote collectively the Riemann curvature and the Yang-Mills field 
strength.

The commutator algebra (\ref{diffa}) implies a generalized Jacobi 
identity
\beq
\sum_{cyclic(MNP)} sign(MNP)  [\Delta_M,[\Delta_N,\Delta_P]] = 0
\eeq
which is the first Bianchi identity for the structure functions 
$\F{MN}{K}$
\beq
\sum_{cyclic(MNP)} sign(MNP) (\Delta_M \F{NP}{K}- \F{MN}{L}\F{LP}{K})=0
\ . 
\eeq
It involves  the sum over the cyclic permutations of $M,\ N,\ P$. If the 
algebra contains fermionic covariant derivatives then there are
additional signs $sign(MNP)$ for each odd permutation of indices of
fermionic covariant derivatives. 

The covariant operations are not defined on arbitrary
polynomials of the jet variables. In particular one cannot realize the 
commutator algebra (\ref{diffa}) on connections, on ghosts
or on auxiliary fields.

To keep the discussion simple we will 
not consider fermionic covariant derivatives in the following. Then the 
commutator algebra (\ref{diffa}) has more specifically the structure
\bea
\ [D_a,D_b] &=& -\T{ab}{c}D_c -\A{F}{ab}{I}\delta_I \quad \mbox{torsion 
and field strength}\\
\ [\delta_I,D_a] &=& -G_{I\,a}^{\phantom{I\,a}b}D_b \quad 
\mbox{representation matrices}\\
\ [\delta_I,\delta_J] &=& \f{IJ}{K}\delta_K \quad \mbox{structure 
constants}\ .
\eea
We will simplify this algebra even more and choose the spin connection 
by the requirement that the torsion vanishes.

The field content $\phi$ of gravitational Yang-Mills theories consists 
of ghosts $C^N$, antighosts $\bar{C}^N$, auxiliary fields $B^N$, gauge 
potentials (connections) $A_m^{\phantom{m}N}\ m=0,\dots,D-1$ and 
elementary tensor
fields $T$. The gauge potentials, ghosts and auxiliary fields are real 
and correspond to
a basis of the covariant operations $\Delta_M$, i.e. there are 
connections, ghosts and
auxiliary fields for translations ( covariant space time derivatives ),
for Lorentz transformations and for isospin transformations. Matter 
fields are tensors and denoted by $T$.
\beq
\label{fields}
\phi = \{C^N, \bar{C^N},B^N,A_m^{\phantom{m}N}, T\}
\eeq
We define the BRS 
transformation on the antighosts and the auxiliary fields by 
\beq
s \bar{C}^N = i B^N \quad sB^N= 0\ .
\eeq
The BRS transformation of tensors is given by  a sum of covariant 
operations with ghosts as coefficients \cite{brandt1}
\footnote{This chapter is nothing but a slightly streamlined version of 
\cite{brandt1}.}
\beq
\label{st}
s T = C^N\Delta_N T\ .
\eeq
Moreover we consider the exterior 
derivative $d=dx^m\partial_m$. We require that the action of partial 
derivatives $\partial_m$ on tensors can be expressed as a
combination of covariant operations with the connections as coefficients
\beq
\label{dten}
d T = dx^m\partial_m T = dx^m A_m^{\phantom{m}N}\Delta_N T = A^N
\Delta_N T\ .
\eeq
If we use the connection one forms
\beq
A^N   = dx^m A_m^{\phantom{m}N}
\eeq
introduced in the last equation then $s$ and $d$ act on tensors in a 
strikingly similar way: $s T$ contains ghosts $C^N$ where $d T$ contains 
(composite) connection one forms $A^N$.

Let us check that (\ref{dten}) is nothing but the usual definition 
of covariant derivatives. We spell out the sum over covariant 
operations and denote the connection $A_m^{\phantom{m}a}$
by  $e_m^{\phantom{m}a}$, the vielbein.
\beq
\partial_m = A_m^{\phantom{m}N} \Delta_M = 
e_m^{\phantom{m}a}D_a + A_m^{\phantom{m}I}\delta_I
\eeq
If the vielbein has an inverse $\A{E}{a}{m}$, which we take for
granted like the rest of the world,
\beq
e_m^{\phantom{m}a}E_a^{\phantom{a}n}=\delta_m^{\phantom{m}n}
\eeq
then we can solve for the covariant space time derivative and obtain the
usual expression
\beq
\label{covder}
D_a = E_a^{\phantom{a}m}(\partial_m - A_m^{\phantom{m}I}\delta_I)\ .
\eeq

We require that $s$ and $d$ anticommute and be nilpotent 
(\ref{anticom}).
This fixes the BRS transformation of the ghosts and  the connection and 
identifies the curvature and field strength. In particular $s^2 = 0 $ 
implies
\beq
0=s^2 T = s (C^N\Delta_N T) = (sC^N)\Delta_nT - C^N s(\Delta_N T)\ .
\eeq
$\Delta_N T$ is a tensor so
\beq
C^N s(\Delta_N T) = C^N C^M \Delta_M \Delta_N T = 
\frac{1}{2}C^N C^M [\Delta_M, \Delta_N ] T\ .
\eeq
The commutator is given by the algebra (\ref{diffa}) and we conclude
\beq
0= (sC^N - \frac{1}{2}C^KC^L\F{LK}{N})\Delta_N T\quad \forall T\ .
\eeq
This means that the operation 
$(sC^N - \frac{1}{2}C^KC^L\F{LK}{N})\Delta_N $ vanishes.
The covariant operations $\Delta_N$ are understood to be linearly 
independent. Therefore $sC^N$ is fixed.
\beq
\label{scn}
sC^N = \frac{1}{2}C^KC^L\F{LK}{N}
\eeq
The BRS transformation of the ghosts is given by a polynomial which is 
quadratic in the ghosts with expansion coefficients given by the 
structure functions $\F{LK}{N}$. 
$s$ transforms the algebra of polynomials generated by  ghosts (not
derivatives of ghosts) and tensors into itself (\ref{st}, \ref{scn}).

The requirement that $s$ and $d$ anticommute fixes the transformation of
the connection.
\bea
\nonumber
0&=&\{s,d\}T=s(A^N\Delta_N T) + d(C^N\Delta_N T)\\
\nonumber
 &=&(sA^N)\Delta_NT - A^NC^M\Delta_M\Delta_NT
+ (dC^N)\Delta_N T - C^NA^M\Delta_M\Delta_N T\\
\nonumber
&=&(sA^N +dC^N - A^KC^L\F{LK}{N})\Delta_N T \quad \forall T
\eea
So we conclude
\beq
sA^N = -dC^N  + A^KC^L\F{LK}{N}
\eeq
for the connection one form $A^N$. For the gauge field
$A_m^{\phantom{m}N}$ we obtain
\footnote{Anticommuting $dx^m$ through $s$ changes the signs.}
\beq
sA_m^{\phantom{m}N} = \partial_mC^N  - A_m^{\phantom{m}K}C^L\F{LK}{N}  
\eeq
The BRS transformation of the connection contains
the characteristic inhomogeneous piece $\partial_mC^N$. The difference 
$\delta \A{A}{m}{N}$ of two connections transforms as a tensor under the
adjoint representation
\beq                                                                                  
s\ \delta A_m^{\phantom{m}N} =  C^L \Delta_L \delta A_m^{\phantom{m}N} 
= C^L\F{KL}{N}\delta A_m^{\phantom{m}K}
\eeq

$d^2=0$ identifies the field strength as curl of the connection.
\bea
\nonumber
0&=&d^2 T= dx^mdx^n\partial_m\partial_n T =
dx^mdx^n\partial_m (\A{A}{n}{N}\Delta_N T)\\
\nonumber
&=& dx^mdx^n \left [ (\partial_m \A{A}{n}{N}) \Delta_N T + 
\A{A}{n}{N} \partial_m (\Delta_N T) \right ] \\
\nonumber
&=& dx^mdx^n \left [ (\partial_m \A{A}{n}{N}) \Delta_N T + 
\A{A}{n}{N} \A{A}{m}{M}\Delta_M\Delta_N T \right ]
\eea
Therefore
\beq
0= \partial_m \A{A}{n}{K} - \partial_n \A{A}{m}{K} +
\A{A}{m}{M}\A{A}{n}{N}\F{MN}{K}
\eeq
We split the summation over $M\,N$, employ the definition
of the vielbein
\bea
\nonumber
0= \partial_m \A{A}{n}{K} - \partial_n \A{A}{m}{K} &+&      
\A{e}{m}{a}\A{e}{n}{b}\F{ab}{K}   +
\A{e}{m}{a}\A{A}{n}{I}\F{aI}{K} \\
\nonumber
&+&
\A{A}{m}{I}\A{e}{n}{a}\F{Ia}{K} +
\A{A}{m}{I}\A{A}{n}{J}\F{IJ}{K}                           
\eea
 and solve for $\A{\cal F}{ab}{K}\ \scriptstyle{K\in (a,[a,b],I)}$.
\bea
\nonumber
\F{ab}{K}  = - \A{E}{a}{m}\A{E}{b}{n} &\left ( \right .&
 \partial_m \A{A}{n}{K} - \partial_n \A{A}{m}{K}
+ \A{e}{m}{c}\A{A}{n}{I}\F{cI}{K}\\
&& \left . +\A{A}{m}{I}\A{e}{n}{c}\F{Ic}{K} +
\A{A}{m}{I}\A{A}{n}{J}\F{IJ}{K}  \right )                          
\eea
The structure functions
\beq
\A{F}{ab}{K}= -\A{\cal F}{ab}{K}
\eeq
are the torsion $\A{T}{ab}{c}$,
\footnote{We require $\A{T}{ab}{c}=0$ which amounts to a choice of
the spin connection.}
 if $K=c$ corresponds to space-time
translations, the Riemann curvature $\A{R}{ab}{cd}$, if $K=[cd]$ 
corresponds to Lorentz transformations, and the Yang-Mills field 
strength $\A{F}{ab}{i}$, if $K=i$ ranges over isospin indices. The 
formula applies, however, also to supergravity, 
which has a more complicated algebra (\ref{diffa}). It allows  in a 
surprisingly simple way to identify the Rarita-Schwinger field strength 
$\A{\Psi}{ab}{\alpha}$ when $K=\alpha$ corresponds to supersymmetry
transformations.

The formulas
\beq
s T = C^N\Delta_N T \quad d T = A^N \Delta_N T
\eeq
for the nilpotent, anticommuting operations $s$ and $d$ not only encrypt 
the basic geometric structures. They allow also
to prove easily that the cohomologies of $s$ and $s+d$ acting on tensors
and ghosts (\underline{not} on connections, derivatives of ghosts, 
auxiliary fields and antighosts ) differ only by a change of variables.
Let us inspect $(s+d)T$.
\beq
\label{stt}
\tilde{s}T = (s+d)T=(C^N+A^N)\Delta_N T = \tilde{C}^N\Delta_N T
\eeq
where
\beq
\tilde{C^N}= C^N + A^N = C^N + dx^m\A{A}{m}{N}
\eeq
The $\tilde{s}$-transformation of tensors is obtained from the 
$s$-transformation by replacing the ghosts $C$ by $\tilde{C}$.

The $\tilde{s}$-transformation of $\tilde{C}$ follows from 
$\tilde{s}^2=0$ and the transformation of tensors (\ref{stt}) by the 
same arguments which
determined $sC$ from $s^2=0$ and from $(\ref{st})$ and led to 
(\ref{scn}). So we obtain
\beq
\label{stc}
\tilde{s}\tilde{C}^N= \frac{1}{2}\tilde{C}^K\tilde{C}^L\F{KL}{N}\ .
\eeq
This is just the tilded version of (\ref{scn}).
Define the map $\rho$ to substitute ghosts $C$ by $\tilde{C}$ in
arbitrary polynomials $P$ of ghosts and tensors.
\beq
P(\tilde{C},T)=\rho \circ P(C,T)
\quad \rho = 
\exp (A\frac{\partial}{\partial C})
\eeq
Taken together (\ref{stt},\ \ref{stc}) and (\ref{st},\ \ref{scn}) imply
\beq
\tilde{s}\circ \rho = \rho \circ s 
\eeq
From this equation one easily concludes the following theorem. 
\begin{theorem}{\quad}\\
\label{sequiv}
Let $s$ be the BRS operation in gravitational theories.
A form $\omega(C,T)$ solves $s\,\omega(C,T)=0$ if and only if
$\omega(\tilde{C},T)$ solves
$\tilde{s}\,\omega(\tilde{C},T)=0$.
\end{theorem}
If we combine this result with theorem (\ref{absts}) then the solutions 
to the descent equations can be found from the cohomology of $s$ if we
can restrict the jet variables to ghosts and tensors. Actually we can 
make this restriction if the base manifold and the 
target space of the fields have trivial topology. This follows because 
the algebra of jet variables is a product of algebras on which
$\tilde{s}$ acts separately. Using
K\"unneth's formula (theorem \ref{kunn}) we can then 
determine nontrivial Lagrange densities and anomaly candidates 
as solutions of $\tilde{s}\omega(\tilde{C},T)=0$ and by
determination of the cohomology of $d$ in the base manifold and of
$\tilde{s}$ in the target manifold.

To establish this result 
we prove the following theorem:
\begin{theorem}{\quad}\\
\label{Asplit}
The algebra $A$ of series in $x^m$ and the fields $\phi 
$ (\ref{fields}) and of polynomials in $dx^m$ and the partial derivatives 
of the fields is a product algebra
\beq
\label{aprod}
A= A_{\tilde{C},T}\times \prod_l A_{u_l,\tilde{s}u_l} 
\eeq
where the variables $u_l$ are given by the following set
\beq
\left ( u_l\right ) = \left (x^m,\A{e}{m}{a}, \A{A}{m}{I},
\partial_{(m_k}\dots \partial_{m_1} \A{A}{{m_0)}}{N},  \bar{C}^N, 
\partial_{m_k}\dots \partial_{m_1} \bar{C}^N \right )
\eeq
for  $k=1,2,\dots$ .
$\tilde{s}$ acts on each factor $A_i$ separately $\tilde{s}\,A_i 
\subset A_i$ .
\end{theorem}
In (\ref{aprod})  the braces around indices denote symmetrization.
The subscript of the algebras denote the generating 
elements e.g.
$A_{\A{e}{m}{a},\tilde{s}\A{e}{m}{a}}$ is the algebra of series in the 
vielbein $\A{e}{m}{a}$ and in $\tilde{s}\A{e}{m}{a}$.
$\tilde{s}$ leaves $A_{u_l,\tilde{s}u_l}$ invariant by construction 
because of $\tilde{s}^2=0$.

To prove the theorem we inspect the variables $u_l$ and $\tilde{s}u_l$ 
to lowest order in the differentials and fields.
\footnote{We do not count powers of the vielbein  $\A{e}{m}{a}$ or its
inverse. Derivatives of the vielbein, however, are counted.}
In lowest order the variables 
$\tilde{s}u_l$ are given by 
\beq
\left (\tilde{s}\, u_l\right ) \approx \left ( dx^m, \partial_m C^a, 
\partial_m C^I, \partial_{m_k}\dots \partial_{m_0}C^N ,
i B^N, 
i \partial_{m_k}\dots \partial_{m_1} B^N 
\right )
\eeq
We recall that to lowest order the covariant derivatives of the 
field strengths are given by 
\beq
\left (T\right )\approx \left (\A{E}{a_k}{m_k}\dots \A{E}{a_0}{m_0}
\partial_{m_k}\dots \partial_{[m_1}\A{A}{m_0]}{N},\ k=1,2,\dots \right 
)\ . \eeq
The brackets denote antisymmetrization of the enclosed indices. 
In linearized order we find all jet variables as linear combinations 
of  the variables $\tilde{C},T, u_l$ and $\tilde{s}u_l $: the 
symmetrized derivatives of the connections belong to
$\left (u_l\right )$, the antisymmetrized derivatives of the connections
belong to the field strengths listed as $T$.
The derivatives of the vielbein are slightly tricky. The symmetrized 
derivatives are contained in
$\partial_{(m_k}\dots \partial_{m_1} \A{A}{{m_0)}}{N}$
for $N=a$. The antisymmetrized derivatives are in one to one 
correspondence to the spin connection $\omega_{k\ [a,b]}$
($\A{A}{k}{I} $ for $I=[ab]$). 
We choose the spin connection $\A{\omega}{ma}{b}$ and a symmetric affine 
connection $\A{\Gamma}{mn}{l}=\A{\Gamma}{nm}{l}$ not
to be elementary variables and determine them from the equations
$D_a \A{e}{n}{b}=0$ and $\A{T}{ab}{c}=0$. This choice does not restrict 
the validity of our investigation because a different choice amounts 
only to the introduction of additional tensor fields.
\beq\partial_m\A{e}{n}{c} - \partial_n\A{e}{m}{c}
= \A{\omega}{ma}{c}\A{e}{n}{a} - \A{\omega}{na}{c}\A{e}{m}{a}
\eeq
\bea
\nonumber
\omega_{k\ [a,b]}&=&\frac{1}{2}\left ( 
\A{E}{a}{m}\eta_{bc}(\partial_k\A{e}{m}{c} - \partial_k\A{e}{m}{c}) -
\A{E}{b}{m}\eta_{ac}(\partial_k\A{e}{m}{c} - \partial_k\A{e}{m}{c}) -
\right . \\
&&\left . -\A{E}{a}{m}\A{E}{b}{n}(\partial_m\A{e}{n}{c} - 
\partial_n\A{e}{m}{c}) \A{e}{k}{d}\eta_{cd}
\right )
\eea
We conclude that the transformation of the jet variables to the 
variables $\left ( \tilde{C},T, u_l,\tilde{s}u_l \right )$ has the
structure
\beq
\phi^{\prime i} = M^i_{\phantom{i}j} \phi^j + O^i(\phi^2)
\eeq
where $M$ is an invertible matrix. 
\beq
\label{phistrich}
\phi^{i} = {M^{-1}}^i_{\phantom{i}j} (\phi^{\prime j} - O^j(\phi^2))
\eeq

Consider an element of the algebra $A$ generated by the jet variables.
We show that it can be written as an element of 
$A_{\tilde{C},T}\times \prod_l A_{u_l,\tilde{s}u_l} $. This holds 
trivially for the variables $x^m$ and $dx^m$ which coincide with $x^m$ 
and $\tilde{s}x^m$. For the remaining variables we neglect in a 
first step all differentials in (\ref{phistrich}).
Concentrate on the terms with the highest derivatives in the expression
for each $\phi^{ i}$. 
The terms $O(\phi^2)$ contain only lower derivatives. Therefore, using 
(\ref{phistrich}), we can
recursively substitute in a polynomial in $\phi$ the highest derivative 
terms by $\phi^{\prime i}$ variables. This changes the expression for 
the lower derivative terms. Then substitute the second highest 
derivative terms. They can be expressed in terms of $\phi^{\prime i}$
with changed terms with third highest derivatives and so on. Therefore 
each polynomial in $\phi$ can be written in terms of $\phi^{\prime i}$.
In a second step we take into account the differentials which come into
play because we use the variables $\tilde{s}u_l$ and therefore 
$O^i(\phi^2)$
contains also the variables $dx^m$ combined also with higher 
derivatives than $\phi^i$. Given an arbitrary
differential form $\omega$ we apply our substitution procedure first to
the zero form. It can be expressed as zero form in the variables
$\phi^{\prime i}$ but the 1-form part has changed. The substitution 
procedure applied to this 1-form part expresses it in terms of
$\phi^{\prime i}$ and changes the 2-form and so on. We iterate the 
substitution  until we reach $D+1$-forms which vanish. Then we have 
expressed the elements of the algebra $A$ of the jet variables in 
terms of the product algebra $A_{\tilde{C},T}\times \prod_l A_{u_l,\tilde{s}u_l} $.
This completes the proof of the theorem.$\Box$

By Knneth's theorem (theorem \ref{kunn}) the cohomology of $\tilde{s}$ 
acting on the algebra $A$ of the jet variables is given by the product 
of the cohomologies of $\tilde{s}$ acting on the
ghost tensor algebra $A_{\tilde{C},T}$ and on the 
algebras $A_{u_l,\tilde{s}u_l}$ 
\beq
\label{cohprod}
H(A,\tilde{s}) = H(A_{\tilde{C},T},\tilde{s})\times \prod_l 
H(A_{u_l,\tilde{s}u_l},\tilde{s})\ .
\eeq

By the Basic Lemma (theorem \ref{basic}) the cohomology of $d$ acting 
on an algebra $A_{x,dx}$ of differential forms $f(x,dx)$ which depend on
generating and independent variables $x$ and $dx$ is given by numbers
$f_0$. Exchanging the names $d$ by $\tilde{s}$ and $x,dx$ by 
$u_l, \tilde{s}u_l$ one can copy the Basic Lemma and conclude that
the cohomology  $H(A_{u_l,\tilde{s}u_l},\tilde{s})$ is given by numbers. 
One can apply this argument if the variables $u_l$ and $\tilde{s}u_l$ 
are independent and not subject to constraints.

Whether the variables $u_l, \tilde{s}u_l$ are subject to constraints is 
a matter of choice of the theory which one considers. This choice 
influences
the cohomology. For example, one could require that two coordinates 
$x^1$ and
$x^2$  satify $(x^1)^2 +(x^2)^2= 1$ because one wants to consider a
theory on a circle. Then the differential
$d(\mbox{arctan}\frac{y}{x})=d\varphi$ is closed 
($d d\varphi=0$) but not exact, because the angle $\varphi$ is not a 
function on the circle, $d\varphi $ is just a misleading notation for a 
one form which is not $d$ of a function $\varphi$. In this 
example the periodic boundary condition
$\varphi \sim \varphi + 2\pi$ gives rise to a nontrivial cohomology
of $d$ acting on $\varphi$ and $d\varphi$. Nontrivial cohomologies also
arise if the fields take values in nontrivial spaces. For example if 
in nonlinear sigma models one requires scalar fields $\phi^i$ to take 
values on a sphere $\sum_{i=1}^n  {\phi^i}^2=1 $ then the volume form 
$d^n \phi$ is nontrivial. More complicated is the case where scalar 
fields are restricted to take values in a general coset $G/H$. Also the 
relation \beq 
\mbox{det}\A{e}{m}{a}\neq 0
\eeq
restricts the vielbeine to take values in the group $GL(D)$ of 
invertible real $D\times D$ matrices. This group has a nontrivial 
cohomology.

For several reasons we choose to neglect the cohomologies coming from a
nontrivial topology of the base manifold with coordinates $x^m$ or the
target space with coordinates $\phi$ or $\A{e}{m}{a}$. 

We have to determine the cohomology of $\tilde{s}$ on the ghost tensor
variables anyhow and start with this problem. To obtain the complete
answer we can determine the cohomology of the base space and the target 
space in a second step which we postpone. 

One can also legitimately argue that perturbation theory replaces fields
by deviations from a ground state and thereby replaces the target space
by its tangent space with a trivial cohomology. 

Canonical quantization does not respect 
inequalites like $x \neq 0$. If there exists a conjugate variable 
$p$ with $[x,p]= -i$ and if the unitary operators $U(y)=e^{iyp}$ exist 
for all real numbers $y$ then the spectrum of $x$ extends over the real 
line including $x=0$. So how does one control the more complicated
inequality $\mbox{det}\A{e}{m}{a}\neq 0$ after quantization?

Whether one accepts these arguments is a matter of choice until the 
physical differences of different choices are calculated and 
tested in nature. We choose to investigate topologically trivial 
base manifolds and target spaces. We combine eq. (\ref{cohprod})
with theorem (\ref{absts} ) and theorem (\ref{sequiv}) and conclude
\begin{theorem}{\quad}\\
If the target space and the base manifold have trivial topology then
the nontrivial solutions of the descent equations  in gravitational 
theories are in one to one correspondence to the nontrivial solutions 
$\omega(C,T)$ of the equation $s\,\omega = 0$. The
relative cohomology (\ref{relcoh1})  is given
by the D-form parts of the forms $\omega(C+A,T)\bmod \tilde{s}\eta$.
\end{theorem}

$\omega$ depends only the ghosts, not on their derivatives. Therefore
the ghost number of $\omega$ is bounded by the number
of translations ghosts and the number of ghosts for spin and for
isospin transformations
$D+\frac{D(D-1)}{2}+\mbox{dim}(G)$. If we take the $D$-form part of
$\omega(C+A,T)$ then $D$ differentials $dx^m$
rather than ghosts have to be picked. Therefore the ghost number of 
nontrivial solutions of the relative cohomology is bounded by
$\frac{D(D-1)}{2}+\mbox{dim}(G)$. This argument, however, does not
apply if there are commuting ghosts for supersymmetry transformations.

From this theorem one can conclude that in an appropriate basis of 
variables anomaly candidates can be chosen such that they contain no
ghosts 
\beq
C^m=C^a\A{E}{a}{m}
\eeq 
of coordinate transformations or in other words that coordinate
transformations are not anomalous.
This result holds if one uses the variables
\beq
\label{chat}
\hat{C}^I=C^I-C^a\A{E}{a}{m}\A{A}{m}{I} 
\qquad \hat{C}^m=C^a\A{E}{a}{m}
\eeq
as ghost fields. This  choice is not very suitable if one wants to split
the algebra of $\tilde{s}$ and therefore we have preferred not to work 
with $\hat{C}^I$. But this choice arises naturally if one enlarges the 
BRS transformation of Yang Mills theories to allow also general 
coordinate transformations. In our formulation the BRS 
transformation is given by
\beq
\nonumber
s T = C^N \Delta_N T = C^a \A{E}{a}{m}(\partial_m - 
\A{A}{m}{I}\delta_I)T + C^I \delta_I T.
\eeq
In the basis of $\hat{C}^m$, $\hat{C}^I$ this is a shift term 
$C^m\partial_m T$ and the BRS transformation of a Yang Mills theory 
\beq
s T = \hat{C}^m\partial_m T + \hat{C}^I \delta_I T\ .
\eeq
The variables $\hat{C}^m$, $\hat{C}^I$ change very simply under the 
substitution of $C$ by $C+A$. 
\beq
\hat{C}^m(C+A)= \hat{C}^m + dx^m 
\quad \hat{C}^I (C+A)=\hat{C}^I
\eeq
If one expresses a form $\omega(C+A,T)$ by ghost variables
$\hat{C}^m$, $\hat{C}^I$ then $\omega$ depends on $dx^m$ only via the 
combination $\hat{C}^m + dx^m$. The $D$ form part $\omega_D$ 
originates from a coefficient function multiplying
\beq
\nonumber
(\hat{C}^1 + dx^1)(\hat{C}^2 + dx^2)\dots (\hat{C}^D + dx^D)=
(dx^1dx^2\dots dx^D + \dots) \ .
\eeq
This coefficient function cannot contain a translation ghost
$C^m=\hat{C}^m$ because $D+1$ factors of translation
ghosts vanish.


\chapter{BRS cohomology on ghosts and tensors}
In the last chapter the problem to determine Lagrange densities 
and
anomaly candidates has been reduced 
to the calculation of the cohomology of $s$ acting
on tensors and ghosts. Let us recall this transformation $s$ explicitly
\footnote{We use a spin connection which makes $\A{T}{ab}{c}$ vanish.}
\bea
\label{defs}
sT&=&(C^aD_a + C^I\delta_I ) T\\
sC^a&=&C^IC^b\A{G}{Ib}{a}\\
sC^I&=& -\frac{1}{2}C^KC^L\A{f}{KL}{I}+\frac{1}{2}C^aC^b\A{F}{ab}{I}\ .
\eea
The BRS transformation 
\beq
\label{splits}
s=s_0+s_1+s_2
\eeq
 consists of a nilpotent part $s_0$
\bea
\label{defs0}
s_0T &=& C^I\delta_I T\\
s_0C^a &=& C^I\A{G}{Ib}{a}C^a\\
s_0 C^I &=&  -\frac{1}{2}C^KC^L\A{f}{KL}{I} \ ,
\eea
which does not increase the number of translation ghosts $C^a$, and of
parts $s_1$
\beq
\label{defs1}
s_1 T =C^a D_a T\quad s_1 C^a = 0 \quad s_1 C^I = 0
\eeq
and $s_2$
\beq
\label{defs2}
s_2 T =0\quad s_2C^a = 0\quad s_2 C^I =\frac{1}{2}C^aC^b\A{F}{ab}{I}\ ,
\eeq
which increase the number of translation ghosts
by 1 and 2. The fact that $s_0^2$ vanishes follows easily if one splits
\beq
\label{splits2}
s^2 = s_0^2 + \{s_0,s_1\} + (\{s_0,s_2\} + s_1^2) + \{s_1,s_2\} + s_2^2 
= 0
\eeq
into pieces which raise the number of translation ghosts by 
0,1,2,3,4. These different pieces vanish separately.

$s_0$ acts on tensors and ghost exactly like the BRS transformation in
Yang Mills theories -- if one interprets the $s_0$ transformation of the
translation ghosts $C^a$ as the BRS transformation of an additional
tensor. 

Let us split each solution $\omega(C,T)$ of $s \omega = 0$ into pieces 
$\omega_n$ which are homogeneous of degree $n$ in translation ghosts 
\beq
\label{omser}
\omega = \omega_{\underline n}+\sum_{n>\underline n}\omega_n + s\eta\ .
\eeq
We call the pieces $\omega_n$ ghosts forms of degree $n$.
Let us concentrate on the ghost form $\omega_{\underline n}$ with 
the lowest degree in $C^a$. It belongs to the $s_0$ cohomology, i.e. it 
satisfies 
\beq
s_0 \omega_{\underline n} = 0 \quad \omega_{\underline n}\bmod 
s_0 \eta_{\underline n} \ .
\eeq
The equation $s_0 \omega_{\underline n} = 0$ is the piece with degree
${\underline n}$ in the equation $s\omega = 0$.
A piece $s_0 \eta_{\underline n}$ can be neglected because it 
is of the form  $s \eta_{\underline n}$ up to pieces with higher degree
in $C^a$ which can be absorbed in a redefined sum
$\sum_{n>\underline n}\omega_n$. Therefore to each element $\omega$ 
of the $s$ cohomology there corresponds an element
$\omega_{\underline n}$ of the $s_0$ cohomology. We  choose $\eta$ such 
that $\underline n$ becomes maximal. Then this correspondence is unique.

To determine $\omega$ we hunt down $\omega_{\underline n}$ and determine
the $s_0$ cohomology. We proceed as in the derivation of the 
Basic Lemma and investigate the anticommutator of $s_0$ with 
other femionic operations. Here we employ the partial derivatives 
with respect to the isospin ghosts $C^I$. These anticommutators coincide
with the generators  $\delta_I$ of isospin transformations
\beq
\label{delant}
\delta_I=\{s_0,\frac{\partial}{\partial C^I}\}
\eeq
which on the ghosts are represented by $G_I$ and the adjoint 
representation
 \beq
\delta_I C^a = \A{G}{Ib}{a}C^b \quad
\delta_I C^J = \A{f}{KI}{J}C^K\ .
\eeq
Eq. (\ref{delant}) is easily verified on the elementary variables 
$C^a,C^I$ and $T$. It extends to arbitrary polynomials because 
both sides of the equation are linear operators with the same product 
rule.

The isospin transformations commute with $s_0$ because each
anticommutator $\{s_0,\delta \}$  of a nilpotent $s_0$ commutes with 
$s_0$ no matter what $\delta$ is (\ref{danti}).
\beq
 [\delta_I,s_0] = 0
\eeq
The representation of the isospin transformations on the algebra of 
ghosts and tensors is completely reducible because the 
isospin transformations belong to a semisimple group or to abelian 
transformations which decompose the algebra into polynomials of definite 
charge and definite dimension. Therefore the following theorem applies.
\begin{theorem}{\quad}\\
\label{sinv}
If the representation of $\delta_I$ is completely reducible then each 
solution of $s_0 \omega~=~0$ is $\delta_I$ invariant up to an 
irrelevant piece.
\beq
s_0 \omega = 0 \Rightarrow  \omega = \omega_{inv}+ s_0 \eta 
\quad \delta_I \omega_{inv}=0
\ .
\eeq
\end{theorem}
The theorem is proven by the following arguments. The 
space \beq 
Z=\{\omega :\ s_0 \omega = 0 \}
\eeq
is mapped by isospin transformations to itself 
$\left ( s_0 (\delta_I \omega ) = \delta_I s_0 \omega = 0 \right )$, 
i.e. $\delta_I Z \subset Z$ . $Z$ contains
the subspace of elements which can be written as  isospin
transformations applied to some other elements $\kappa^I \in Z$
\beq
Z_{\delta}= \{\omega \in Z:\ \omega = \delta_I ( \kappa^I )\quad
s_0\kappa^I =0 \}\ .
\eeq
$Z_\delta$ is mapped by isospin transformations to
itself. A second invariant subspace is given by
$Z_{inv}$,
the subspace of $\delta_I$ invariant elements
\beq
Z_{inv}=\{\omega \in Z:\  \delta_I \omega = 0\ \}\ .
\eeq 
If the representation of $\delta_I$ is completely reducible then
the space $Z$ is spanned by $Z_{inv}\oplus Z_\delta\oplus Z_{comp}$ 
with a complement $Z_{comp}$ which is also mapped to itself. This
complement, however,  
contains only $\omega=0$ because if there were a nonvanishing element 
$\omega \in Z_{comp}$ it would
not be invariant because it is not  from $Z_{inv}$. $\omega$ would be 
mapped to $\delta_I \omega \in Z_{\delta}$ and $Z_{comp}$ would not be 
an invariant subspace.
\beq
Z = Z_{inv}\oplus Z_{\delta}
\eeq
Each  $\omega $ which satisfies $s_0 \omega = 0 $ can therefore be 
decomposed as
\beq
\omega = \omega_{inv}+  \delta_I  \kappa^I \quad s_0 \kappa^I = 0\ .
\eeq
We replace $\delta_I$ by $\{s_0,\frac{\partial}{\partial C^I}\}$ 
(\ref{delant}), use $s_0 \kappa^I = 0$  and verify the theorem.
\beq
\omega = \omega_{inv} + s_0 \eta \quad \eta= 
\frac{\partial}{\partial C^I}\kappa^I
\eeq

The theorem restricts nontrivial solutions to $s_0 \omega = 0$ to spin 
and isospin invariant combinations.

We can exploit this theorem a second 
time and conclude that the translation ghosts $C^a$ and the tensors $T$ 
occur only in invariant combinations and that the ghosts $C^I$ 
of spin and isospin transformations couple separately to invariants. 
This follows from the peculiar form of $s_0$ which is given by 
$C^I\delta_I$ if it acts on translation ghosts and  tensors and 
by $\frac{1}{2} C^I\delta_I $ if it acts on the ghosts $C^I$ of spin and
isospin transformations.
\beq
s_0=C^I\delta_I - s_c
\eeq
$s_c$  transforms only spin and isospin ghosts
\beq
s_c T = 0 \quad s_c C^a = 0 \quad 
s_c C^I = -\frac{1}{2}C^KC^L\A{f}{KL}{I}\ .
\eeq
The equations
$s_0 \omega_{inv}= 0$ and $\delta_I \omega_{inv}=0$ imply
\beq
s_c \omega_{inv}=0 \ .
\eeq
The anticommutator
\beq
\delta_{\scriptscriptstyle{C}^I}=\{s_c,\frac{\partial}{\partial
C^I}\}
\eeq
generates the adjoint transformations of the spin and isospin ghosts. 
\beq
\quad \delta_{\scriptscriptstyle{C}^I} C^J = \A{f}{KI}{J}C^K
\quad \delta_{\scriptscriptstyle{C}^I} T=0
\quad \delta_{\scriptscriptstyle{C}^I} C^a=0
\eeq
It can be used to express $s_c$ in the forms
\beq
\label{scdelta}
s_c=\frac{1}{2}C^I\delta_{\scriptscriptstyle{C}^I}=
\frac{1}{2}\delta_{\scriptscriptstyle{C}^I}C^I
\eeq
which are both valid because $\A{f}{IJ}{I}=0$ in Lie algebras 
which consist of simple and abelian factors.

By  theorem (\ref{sinv}) one can conclude from $s_c\omega_{inv}= 0$ 
that $\omega_{inv}$ consists of a part $\omega_{inv|inv}$
which is invariant under $\delta_{\scriptscriptstyle{C}^I}$ 
and a piece $(s_c \eta)_{inv}$ which is also $s_0$ exact 
because $(s_c\eta)_{inv}$ is $\delta_I$ invariant.
Therefore $\omega$ is of the form
$$
\omega = f(\theta_\alpha (C^I),I_\tau (C^a,T)) + s_0 \eta.
$$
where $\theta_\alpha(C^I)$ and $I_\tau(C^a,T)$ are invariant functions.
A contribution $s_c\eta$ to $\theta_\alpha$  changes $f$ only by 
an irrelevant piece because $s_c \eta\,I(C^a,T)=s_0(\eta\, I)$.
We can therefore state:

\begin{theorem}{\quad}\\
\label{s0coh}
An element $\omega$ of the algebra of ghosts and tensors
satisfies $s_0 \omega = 0$ if and only if it is of the form
\beq
\omega = f(\theta_\alpha (C^I),I_\tau (C^a,T)) + s_0 \eta
\eeq
where $I_\tau(C^a,T)$ are invariant functions and where the invariant 
functions $\theta_\alpha(C^I)\ \alpha=1,\dots,r$ generate the 
Lie algebra cohomology
\beq
\label{liecoh}
s_c \Theta (C^I) = 0 \Leftrightarrow
 \Theta(C^I) = \Phi(\theta_1(C),\dots ,\theta_r(C)) +s_c \eta (C^I) \ .
\eeq
$\omega$ is trivial if and only if $f$ vanishes.
\end{theorem}

The solutions of $s_c\Theta(C)=0$ are given by the  
$\delta_{\scriptscriptstyle{C}^I}$ invariant polynomials $\Theta 
(C^I)$. Obviously theses invariant polynomials  satisfy $s_c \Theta = 0$ 
and they are nontrivial because all trivial solutions $s_c\eta$ are
contained in $Z_\delta$ as eq.(\ref{scdelta}) shows. $Z_\delta$ contains
no invariant elements because $Z=Z_{inv}\oplus Z_\delta$. 
\footnote{By the same argument one shows that $f$ is nontrivial.}

The space of invariant polynomials can be determined separately for each
factor of the Lie algebra. The general solution for the product algebra
can then be obtained with Knneth's formula (theorem \ref{kunn}).

The following results for simple Lie algebras can be found in the 
mathematical literature \cite{greub3} or in translations into a language 
which a (german) physicist is used to \cite{juen}. For a simple Lie 
algebra $G$
the dimension of the space of invariant polynomials $\Theta(C)$ is $2^r$
where $r$ is the rank of $G$. These invariant polynomials are generated
by $r$ primitive polynomials $\theta_\alpha(C), \ \alpha=1,\dots ,r$
which cannot be written as a sum of products of other invariant
polynomials. They have odd ghost number
gh$(\theta_\alpha(C))=2m(\alpha)-1$ and therefore are fermionic. They
can be obtained from traces of suitable matrices $M_i$
which represent the Lie algebra and are given with a suitable 
normalization by
 \beq
\theta_\alpha (C)=
\frac{(-)^{m-1}m!(m-1)!}{(2m-1)!} 
tr(C^i M_i)^{2m -1}
\quad m=m(\alpha) \quad \alpha=1,\dots,r\ .
\eeq
The number $m(\alpha)$ is the degree of homogeneity of the 
corresponding Casimir invariant $I_\alpha(X)$
\beq
I_\alpha(X)=tr (X^i M_i)^{m(\alpha)}\ .
\eeq
These Casimir invariants generate all invariant functions of a 
set of commuting variables $X^i$ which transform as an 
irreducible multiplet under the adjoint representation.

The degrees $m(\alpha)$ for the classical Lie algebras are given by
\beq
\label{mlist}
\begin{array}{l l l l l}
SU(n)&A_{n-1}&m(\alpha)=\alpha+1&\alpha=1,\dots,n-1\\
SO(2n+1)&B_n&m(\alpha)=2\alpha&\alpha=1,\dots,n\\
SP(2n)&C_n&m(\alpha)=2\alpha& \alpha=1,\dots,n\\
SO(2n)&D_n&m(\alpha)=2\alpha&\alpha=1,\dots,n-1&m(n)=n
\end {array}
\eeq

With the exception of the last primitive element of $SO(2n)$
the matrices $M_i$ are the defining representation of the classical
Lie algebras.
The last primitive element $\theta_n$ and the last Casimir invariant 
$I_n$ of $SO(2n)$ are constructed from the spin representation 
$\Gamma_i$.
Up to normalization they are given by
$$
\theta_n \sim \varepsilon_{a_1b_1\dots a_nb_n}(C^2)^{a_1b_2}
\dots (C^2)^{a_{n-1}b_{n-1}}C^{a_nb_n}\quad 
I_n \sim \varepsilon_{a_1b_1\dots a_nb_n}X^{a_1b_2}\dots X^{a_nb_n}\ .
$$
If $n$ is even then the primitive element $\theta_n$ of $SO(2n)$ is 
degenerate in ghost number with $\theta_{\frac{n}{2}}$.

The primitive elements for the exceptional simple Lie algebras 
$G_2,$ $F_4,$ $E_6,$ $ E_7,$ $ E_8$ can also be found in the literature
\cite{oraif}. 
Their explicit form is not important for our purpose. In each case the
Casimir invariant with lowest degree $m$  is quadratic ($m=2$).

For a one dimensional abelian Lie algebra the ghost $C$ is invariant
under the adjoint transformation. It generates the invariant 
polynomials $\Theta(C) = a + b C$ which span a $2^r$ dimensional
space where $r=1$ is the rank of the abelian Lie algebra. The generator
$\theta$ of this algebra of invariant polynomials has odd ghost number
gh$(C)=2m-1$ with $m=1$.
\beq
\theta(C) =  C 
\eeq
The Casimir invariant $I$ of the one dimensional, trivial 
adjoint representation acting on a bosonic variable $X$ is homogeneous
of degree $m=1$ in $X$ and is simply given by $X$ itself.
\beq
I(X)= X\ .
\eeq

If the Lie algebra is a product of simple and abelian factors then
the list of primitive elements $\theta_\alpha$ and the
list of the Casimir invariants $I_\alpha$
are the union of the respective lists of the factors of the Lie 
algebra.

Polynomials of $r$ anticommuting variables $\theta_\alpha$ span
a $2^r$ dimensional space, which theoretical physicists would call a
superspace. The statement that the primitive elements $\theta_\alpha(C)$
span the space of $\delta_I$ invariant polynomials in the anticommuting
ghosts
\beq
\delta_I \Theta(C) = 0 \Rightarrow 
\Theta(C) = \Phi(\theta_1(C),\dots ,\theta_r(C))
\eeq
asserts that the Lie algebra cohomology is given by 
\beq
s_c \Theta(C)= 0 \Leftrightarrow 
\Theta(C) =\Phi(\theta_1(C),\dots,\theta_r(C)) + s_c \eta\ .
\eeq
Because the space of these invariant functions is $2^r$ 
dimensional  there are no algebraic relations among the functions
$\theta_\alpha(C)$ apart from the anticommutation relations which result
from their odd ghost number.
\beq
\Theta(C)=\Phi(\theta_1(C),\dots ,\theta_r(C)) = 0
\Leftrightarrow \Phi(\theta_1,\dots ,\theta_r) = 0
\eeq

The Casimir invariants $I_\alpha(X)$ generate the space of
$\delta_I$ invariant polynomials in commuting variables $X$ which
transform under the adjoint representation
\beq
\delta_I P(X) = 0 \Rightarrow 
P(X)=f(I_1(X),\dots ,I_r(X))=0\ .
\eeq

There is no algebraic relation among the Casimir
invariants $I_\alpha(X)$ up to the fact that the $I_\alpha$ commute 
\cite{greub3}.
\beq
P(X)=f(I_1(X),\dots ,I_r(X))=0 \Leftrightarrow f(I_1,\dots ,I_r)=0
\eeq

Theorem (\ref{s0coh}) describes all  solutions
$\omega_{\underline n}$ of the equation $s_0 \omega_{\underline n}= 0$.
This equation is the part of $s \omega = 0$ with lowest degree in 
the translation ghosts. In degree ${\underline n}+1$ the equation $s \omega = 0$
imposes the restriction
\beq
s_1 \omega_{\underline n}+ s_0 \omega_{{\underline n }+1 }=0\ .
\eeq
We choose  
$\omega_{\underline  n}=f(\theta_\alpha(C),I_\tau(C^a,T))$.
Then $s_1 \omega_{\underline n}$ is $\delta_I$
invariant and not of the form $s_0 \eta$ because $s_1$ (\ref{defs1})
\beq
s_1 T = C^aD_a T\quad s_1 C^a = 0 \quad s_1 C^I = 0
\eeq
maps invariant functions
$I_\tau$ of tensors and translation ghosts to invariant functions. 
Therefore $s_1 \omega_{\underline n}$ has to vanish because it is not 
of the form $s_0 \eta$.

We can require more restrictively that $\omega_{\underline n}$ is an 
element of the $s_1$ cohomology after we consider the following 
argument. A contribution to $\omega_{\underline n}$ of the form 
$s_1\eta(\theta_\alpha,I_\tau)$ can be written as $s \eta - s_2 
\eta $ because $s_0 \eta$ vanishes ($\eta$ is $\delta_I$ invariant).
$s \eta$ changes $\omega=\omega_{\underline n}+ \dots$ only by an
irrelevant piece. $s_2 \eta $ can be absorbed in the parts $\dots$
with higher ghost degree. Therefore we can neglect contributions
$s_1\eta(\theta_\alpha,I_\tau)$ to $\omega_{\underline n}$. 

\beq
s_1 \omega_{\underline n} = 0 \quad \omega_{\underline n} \bmod s_1 
\eta_{inv}
\eeq

The operation $s_1$ acting on {\it invariant} functions is nilpotent
because (\ref{defs0},\ref{defs2},\ref{splits2})
\beq
\label{s1coh}
s_1^2+ \{s_0,s_2\}= 0 = s_1^2 + F^I\hat{ \delta}_I
\eeq
where $F^I$ is the ghost two form
\beq
F^I=\frac{1}{2}C^aC^b\A{F}{ab}{I}                            
\eeq
and $\hat{ \delta}_I $ generates the adjoint transformation of
translation ghosts and tensors
\beq 
\hat{ \delta}_I T = \delta_I T\quad \hat{ \delta}_I C^a =
\delta_I C^a \quad \hat{ \delta}_I C^J = 0\ .
\eeq

$s_1$ is the covariant exterior derivative $D=dx^mD_m$ in disguise.
It does not differentiate the translation ghosts, the relation 
$s_1(C^a) = 0$ corresponds to the relation $d (dx^m) = 0$. An invariant
ghost form of degree $l={\underline n}$ is given by
\beq
\omega(C,T)=\frac{1}{l!}C^{a_1}\dots C^{a_l}\omega_{a_1 \dots a_l}(T)
\eeq
where the components $\omega_{a_1 \dots a_l}$ belong to
an isospin invariant Lorentz tensor which transforms as indicated by 
the index picture. $s_1$ acts on $\omega$ (\ref{defs1}) by
\beq
s_1 \omega=\frac{1}{(l+1)!}C^{a_1}\dots C^{a_{l+1}}
\sum_{cyclic(1,2,\dots,l+1)}sign(cyclic)D_{a_{1}}\omega_{a_2 \dots 
a_{l+1}}\ .
\eeq 
If we convert the index picture from Lorentz indices to space time 
indices by help of the vielbein $\A{e}{m}{a}$ and its inverse 
$\A{E}{a}{m}$ and define the space time covariant derivative $D_m$ and 
the ghosts $C^m$ by
\beq
D_{a_0}\omega_{a_1\dots a_l }=
\A{E}{a_0}{m_0}\A{E}{a_1}{m_1}\dots\A{E}{a_l}{m_l}
D_{m_0}\omega_{m_1\dots m_l} \quad C^m = C^a\A{E}{a}{m}
\eeq
then $s_1$ acts on forms $\omega$ 
\beq
\label{omegaghost}
\omega(C,T)=\frac{1}{l!}C^{m_1}\dots C^{m_l}\omega_{m_1 \dots m_l}(T)
\eeq
in the same way as the exterior covariant derivative $D=dx^mD_m$. 
Only the name of the differential $dx^m$ is changed to $C^m$.

\beq
s_1 \omega=\frac{1}{(l+1)!}C^{m_1}\dots C^{m_{l+1}}
\sum_{cyclic(1,2,\dots,l+1)}sign(cyclic)D_{m_{1}}\omega_{m_2 \dots 
m_{l+1}}\ .
\eeq 

$s_1$ simplifies on $\delta_I$ invariant forms even more because one can 
neglect the isospin transformations in the covariant derivatives
$D_a=\A{e}{a}{m}(\partial_m -\A{A}{m}{I}\delta_I)$. The
spin connection $\A{\omega}{ma}{b}$ in the covariant derivative is 
exchanged for the symmetric Christoffel symbol
\beq
\A{\Gamma}{mn}{k}=
\frac{1}{2}g^{kl}\left ( \partial_m g_{nl} +\partial_n g_{ml}-\partial_l
g_{mn}\right )\quad \A{\Gamma}{mn}{k}=\A{\Gamma}{nm}{k}\quad
g_{mn}=\A{e}{m}{a}\A{e}{n}{b}\eta_{ab}
\eeq
if the Lorentz vector indices $a,b,\dots$ are traded for tangent space
indices $m,n,\dots$. The contributions of these Christoffel symbols 
vanish if $s_1$ is applied to an invariant form because all tangent 
space indices are contracted with anticommuting ghosts $C^m$, e.g.
\bea
\nonumber
s_1 C^n\omega_n &=& s_1 C^b\omega_b =-C^bC^aD_a\omega_b =
C^mC^nD_m\omega_n\\
\nonumber
&=& C^mC^n(\partial_m\omega_n -\A{\Gamma}{mn}{l}\omega_l)
= C^mC^n\partial_m\omega_n 
\eea
Therefore $s_1$ acts on invariant ghost forms is the same way as the 
exterior derivative $d=dx^m\partial_m$ acts on differential forms.

The cohomology of $d$ acting on the jet variables is given by the 
Algebraic Poincar\'e Lemma (theorem (\ref{algebraic})). This lemma, 
however, does not apply here because among the tensors there are the 
field strengths on which the derivatives do not act freely, i.e. with no
constraint apart from the fact that they commute,  but subject to 
the Bianchi identities 
\beq
\sum_{cyclic}D_aF_{bc}= 0\ .
\eeq
These constraints on the action of the derivatives change the cohomology
of $d$. It is given by  the Covariant Poincar\'e Lemma \cite{kovPoinc}
\begin{theorem}{Linearized Covariant Poincar\'e Lemma}\\
Consider functions ${\cal L}$ and differential forms $\omega$ and 
$\eta$ which depend on linearized field strengths 
$\A{F}{mn}{i}=\partial_m\A{A}{n}{i}-\partial_n\A{A}{m}{i}$ and 
their derivatives, which are restricted by 
$\sum_{cyclic}\partial_k\A{F}{mn}{i}= 0$,
and on other fields $\psi$ and their derivatives. If $\omega$ satisfies
$d\omega=0$ then it can be written as a sum of a volume 
form ${\cal L}d^Dx$ and a polynomial  $P(F)$ in the field strength two 
forms $F^i=\frac{1}{2}dx^mdx^n\A{F}{mn}{i}$ and 
an exact form $d\eta$. 
\bea \nonumber &&d\ \omega\left (dx^m,
F_{mn},\partial_{(k}F_{m)n},\dots,\psi,\partial_k\psi,\dots,\partial_{(k}\dots 
\partial_{l)}\psi \right ) = 0
\Leftrightarrow\\
&&\omega = {\cal L}(F_{mn}, \partial_{(k}F_{mn)},\dots ,\psi, 
\partial_k\psi,\dots) d^Dx + P(F) + d\eta
\eea
The Lagrange density ${\cal L}d^Dx$ cannot be written as $P(F)+d\eta$ if
its Euler derivatives (\ref{eulder})
with respect to $\psi$ and $\A{A}{n}{i}$ do not vanish 
\beq
\frac{\hat{\partial} \cal L}{\hat{\partial} \psi }\neq 0
\quad \mbox{ or }\quad
\partial_m \frac{\hat{\partial} \cal
L}{\hat{\partial} \A{F}{mn}{i} }\neq 0 \ .
\eeq
\end{theorem}
A nonvanishing polynomial $P(F)$ cannot be written as $d$ of a form 
$\eta$ which depends on field strengths and fields $\psi$ and their 
derivatives because $\eta$ would have to contain at least one connection
$\A{A}{m}{i}$ without derivative.

The theorem can be extended to cover Lorentz and isospin invariant 
Lagrange densities depending on the (nonlinear) field strengths, other 
tensors and and their covariant derivatives.
 
\begin{theorem}{Covariant Poincar\'e Lemma}\\                                    
\label{covpointheo}
Consider $\delta_I$ invariant functions ${\cal L}$ and differential 
forms $\omega$ and $\eta$ which depend on field strengths  
$\A{F}{mn}{I}$ and their covariant derivatives, which are restricted by                                               
$\sum_{cyclic}D_k\A{F}{mn}{I}= 0$, and on other fields $\psi$ and their
covariant derivatives. If $\omega$ satisfies $d \omega=0$ then it
can be written as a sum of a volume                                     
form ${\cal L}d^Dx$ and an invariant polynomial  $P(F)$ in the 
field strength two forms $F^I=\frac{1}{2}dx^mdx^n\A{F}{mn}{I}$ and                                          
an exact form $d\eta$.                                                                   
\bea \nonumber &&d\ \omega\left (dx^m,                                                   
F_{mn},D_{(k}F_{m)n},\dots,\psi,D_k\psi,\dots,D_{(k
}\dots D_{l)}\psi \right ) = 0                                                         
\Leftrightarrow\\                                                                        
&&\omega = {\cal L}(F_{mn}, D_{(k}F_{m)n},\dots ,\psi,
\dots) d^Dx + P(F) + d\eta                                                 
\eea                                                                                      
The Lagrange density ${\cal L}d^Dx$ cannot be written as 
$P(F)+d\eta$ if
its Euler derivatives                                                                     
with respect to $\psi$ and $\A{A}{n}{I}$ do not vanish                                    
\beq                                                                                      
\frac{\hat{\partial} \cal L}{\hat{\partial} \psi }\neq 0                                  
\quad \mbox{ or }\quad                                                                    
D_m \frac{\hat{\partial} \cal                                                      
L}{\hat{\partial} \A{F}{mn}{I} }\neq 0 \ .                                                
\eeq                                                                                      
\end{theorem}                                                                            

We call the invariant polynomials $P(F^I)$ Chern forms. They are 
polynomials in commuting variables, the field strength two forms $F^I$
which transform as an adjoint representation of the Lie algebra. These
invariant polynomials are generated by the elementary Casimir invariants
$I_\alpha(F^I)$. The Chern forms enlarge the cohomology of the exterior
derivative if it acts on tensors rather than on jet variables. They 
comprise all topological densities which one can construct from
connections for the following reason. If a functional is to contain only
topological information it's value must not change under continuous 
deformation of the fields. Therefore it has to be gauge invariant and
invariant under general coordinate transformations. If it is a local
functional it is the integral over a density which satisfies the descent
equation and which can be obtained from a solution to $s\omega=0$. If
this density belongs to a functional which contains only topological
information then the value of this functional must not change even under
arbitrary differentiable variations of the fields, i.e. its Euler
derivatives with respect to the fields must vanish. Therefore the
integrand must be a total derivative in the space of jet variables. But 
it must not be a total derivative in the space of tensor variables 
because then it would be constant and contain no information at all. 
Therefore, by theorem (\ref{covpointheo}), all topological densities
which one can construct from connections are given by Chern polynomials
in the field strength two form.

Theorem (\ref{covpointheo}) describes also the cohomology of $s_1$ 
acting on invariant ghost forms because $s_1$ acts on invariant ghost
forms (\ref{omegaghost}) exactly like the exterior derivative $d$ acts
on differential forms. We have to allow, however, for the additional 
variables $\theta_\alpha(C)$ in $\omega_{\underline n}$. They 
generate a second, trivial algebra $A_2$ and can be taken into account
by K\"unneth's theorem (theorem (\ref{kunn})). If we neglect the trivial
part $s_1\eta_{inv}$ then the solution to (\ref{s1coh}) is given by
\beq
\label{sol1}
\omega_{\underline n}=
{\cal L}(\theta_\alpha(C),T)C^1C^2\dots C^D + 
P(\theta_\alpha(C),I_\alpha(F))
\eeq
The $\delta_I$ invariant Lagrange ghost density satisfies already the 
complete equation $s\omega(C,T)=0$ because it is a $D$ ghost form. The 
solution to $\tilde{s}\tilde{\omega} = 0$ is given by 
$\tilde{\omega}=\omega(C+A,T)$ and the
Lagrange density and the anomaly candidates are given by the part of 
$\tilde{\omega}$ with $d^Dx$. The coordinate differentials come from
$C^a+dx^m\A{e}{m}{a}$
\footnote{We can use the ghosts variables $C$ or $\hat{C}$ (\ref{chat}).
The expessions remain unchanged because they are multiplied by $D$ 
translation ghosts }.
If one picks the $D$ form part then one gets \beq
dx^{m_1}\dots dx^{m_D}\A{e}{m_1}{1} \dots \A{e}{m_D}{D}=
\det (\A{e}{m}{a})d^Dx \quad   \det (\A{e}{m}{a})=:\sqrt{g}
\eeq
Therefore the solutions to the descent equations of Lagrange type are 
given by
\beq
\omega_D={\cal L}(\theta_\alpha(C),T)\sqrt{g}d^Dx\ .
\eeq
They are constructed in the well known manner from tensors $T$, 
including fields strengths and covariant derivatives of tensors, which
are combined to a Lorentz invariant and isospin invariant Lagrange
function. This composite scalar field is multiplied by the density
$\sqrt{g}$. Integrands of local gauge invariant actions are obtained
from this formula by restricting $\omega_D$ to vanishing ghost number. 
Then the variables $\theta_\alpha(C)$ do not occur. We indicate the 
ghost number by a superscript and have
\beq
\omega_D^0={\cal L}(T)\sqrt{g}d^Dx\ .
\eeq
Integrands of anomaly candidates are obtained by choosing $D$ forms 
with ghost number $1$. Only abelian factors of the Lie algebra allow for
such anomaly candidates because the primitive invariants $\theta_\alpha$
for nonabelian factors have at  least ghost number $3$.
\beq
\omega_D^1=\sum_i C^i{\cal L}_i(T)\sqrt{g}d^Dx\ .
\eeq
The sum ranges over all abelian factors of the gauge group.
Anomalies of this form actually occur as trace anomalies or $\beta$ 
functions if the isospin algebra contains dilatations.

This completes the discussion of Lagrange densities and anomaly 
candidates of the coming from the first term in (\ref{sol1}).


\chapter{Chiral anomalies}
It remains to investigate solutions which correspond to
\beq
\omega_{\underline n}=P(\theta_\alpha(C),I_\alpha(F))\ .
\eeq
Ghosts $C^I$ for spin and isospin transformations and ghost forms $F^I$ 
generate a subalgebra which is invariant under $s$ and takes a 
particularly simple form if expressed in terms of matrices $C=C^IM_I$ 
and $F=F^IM_I$ which represent the Lie algebra. For nearly all algebraic
operations it is irrelevant that $F$ is a composite field. The
transformation of $C$  (\ref{defs}) can be read as definition of
$F=sC + C^2$ and $s^2=0$ determines the transformation of $F$ which is 
given by the adjoint transformation. One calculates
$$sF=sC\,C - C\,sC=(F-C^2)C-C(F-C^2)$$
\beq
sC=-C^2+F \quad sF=FC-CF\ .
\eeq
If one changes the notation and replaces $-s$ by $d=dx^m\partial_m$ and
$-C$ by $A=dx^m\A{A}{m}{I}M_I$ then the same equations are the 
definition
of the field strengths in Yang Mills theories and their Bianchi
identities. The equations are valid whether the anticommuting variables
$C$ and the nilpotent operation $s$ are composite or not.
\footnote{This does not mean that there are no differences at all. For 
example the one form matrices $A$ satisfy $A^{D+1}=0.$}

The Chern polynomials $I_\alpha$ satisfy $sI_\alpha=0$ because they are 
invariant under adjoint transformations. All $I_\alpha$ are trivial i.e.
of the form $sq_\alpha$. To show this explicitly we
define a one parameter deformation $F(t)$ of $F$
\beq
F(t)=tF+(t^2-t)C^2=t\,sC+t^2C^2\quad F(0)=0 \quad F(1)=F
\eeq
which allows to switch on $F$.

All invariants $I_\alpha$ can be written as $tr(F)^{m(\alpha)}$ with 
suitable representations $M_I$. We rewrite $tr(F)^m$ in an artificially
more complicated form $$
tr (F)^m = \int_o^1 dt \frac{d}{dt}tr (F(t))^m=m\int_0^1 dt 
\,tr \left( (sC+2tC^2)F(t)^{m-1}\right)\ .
$$
The integrand coincides with
\bea
\nonumber
s\,tr\,(C F(t)^{n-1})&=&tr\,((sC)F(t)^{n-1}-tC[F(t)^{n-1},C])\\
\nonumber
&=&tr(sCF(t)^{n-1}+ 2tC^2F(t)^{n-1})\ .
\eea
The Chern form $I_\alpha$ is the $s$ transformation of the Chern Simons
form $q_\alpha$, these forms generate a subalgebra.
\beq
\label{cherncoh}
sq_\alpha=I_\alpha \quad sI_\alpha=0
\eeq
\beq
q_\alpha=m\int_0^1dt\  tr \left( C\left (tF+(t^2-t)C^2\right )^{m-1}
\right ) \quad m=m(\alpha) 
\eeq
The $t$-integral gives the combinatorial coefficients of the Chern 
Simons form.
\beq
\label{csimons}
q_\alpha(C,F)=\sum\limits_{l=0}^{m-1}\frac{(-)^lm!(m-1)!}
{(m+l)!(m-l-1)!} tr_{sym}\left(C (C^2)^l  (F)^{m-l-1} \right)
\eeq
It involves the traces of completely smmetrized products of the $l$ 
factors $C^2$, the $m-l-1$ factors $F$ and the factor $C$. The part with
$l=m-1$ has form degree $0$ and ghost number $2m-1$ and agrees
with $\theta_\alpha$
\beq
q_\alpha(C,0)=\frac{(-)^{m-1}m!(m-1)!}{(2m-1)!}tr 
C^{2m-1}=\theta_\alpha(C)\ . \eeq
Each polynomial 
$\omega_{\underline n}= P(\theta_\alpha(C),I_\alpha(F))$ defines 
naturally a form 
\beq
\omega(C,F)=P(q_\alpha(C,F),I_\alpha(F))
\eeq
which coincides with 
$\omega_{\underline n}$ in lowest form degree.
\beq
\omega(q_\alpha(C,F),I_\alpha(F))=
\omega_{\underline n}(\theta_\alpha(C),I_\alpha(F)) + \dots
\eeq
On such forms $s$ acts in a very simple way.
\beq
\label{sqi}
\left . s \omega = I_\alpha \frac{\partial}{\partial q_\alpha} 
P(q_\alpha,I_\alpha) \right ._{|q_\alpha(C,F),I_\alpha(F)}
\eeq

We define the level $m$ of a monomial 
$M=c \prod I_{\alpha_i}q_{\alpha_j}$ as the lowest degree 
$m(\alpha)$ of the variables $I_{\alpha_i}$ and $q_{\alpha_j}$ which 
actually occur in $M$
\beq
m(M) = \min \ \{m(\alpha):\ 
I_\alpha\frac{\partial}{\partial I_\alpha}_{|no\  sum}M\neq 0 
\ \mbox{ or }\ 
q_\alpha\frac{\partial}{\partial q_\alpha}_{|no\  sum}M \neq 0 \}\ .
\eeq
This definition decomposes the polynomial $P$ naturally into polynomials
$P_m$ with definite level (which consist of monomials with level $m$)
\beq
P =\sum_m P_m + const\ .
\eeq
$s$ does not mix levels (\ref{sqi}). Therefore we can consider the 
each $P_m$ separately.

We decompose the space ${\cal P}_m$ of polynomials $P_m$   la Hodge 
(\ref{hodge}) with the operators \footnote{Hopefully the $s_m$ are not 
confused with $s_0,s_1,s_2$ defined in 
(\ref{defs0}$-$\ref{defs2})} 
\beq
s_m=
\sum_{m(\alpha)=m}I_\alpha\frac{\partial}{\partial q_\alpha}
\qquad
r_m=
\sum_{m(\alpha)=m}q_\alpha\frac{\partial}{\partial I_\alpha}
\eeq
into ${\cal S}_m=s_m{\cal P}_m$ and ${\cal R}_m=r_m{\cal P}_m$
\beq
{\cal P}_m={\cal S}_m \oplus {\cal R}_m
\eeq
and write $P_m=S_m + R_m$ as a $s_m$ exact piece $S_m$ and a
$r_m$ exact piece $R_m$
\footnote{Unluckily the alphabet is a small set. Do not interpret $A$ 
and $B$ in these formulas as connection one form or as auxiliary 
field.}
\beq P_m=S_m + R_m \quad S_m = s_mA \quad R_m= r_m B 
\eeq
Without loss of generality we can taken $A$ from ${\cal R}_m$ and 
$B$ from ${\cal S}_m$.

The piece $S_m$ can be rewritten as a trivial contribution to 
$\omega$ and a  part which lies in ${\cal R}_m$ because 
$A \in {\cal R}_m$ 
\beq
\label{trivsm}
s_mA= sA - \sum_{m^\prime \ge m+1}s_{m^\prime}A = sA + 
A^\prime \quad A^\prime \in {\cal R}_m
\eeq
Eq. (\ref{trivsm}) holds because $s=\sum_m s_m$ and $s_{m^\prime}A=0$ 
for ${m^\prime}<m$. Therefore we can restrict $P_m$ to the $r_m$ exact 
part.
\beq
P_m=r_m B \quad  B \in {\cal S}_m
\eeq
Such a polynomial $P_m(q,I)$, however, cannot be made to satisfy $sP_m= 
0$ \beq
sP_m=s_mr_m B+\sum_{m^\prime \ge m+1}s_{m^\prime}r_mB
=N_m B + \sum_{m^\prime \ge m+1}s_{m^\prime}r_mB\ .
\eeq
The pieces $s_mr_mB$ and the sum have to vanish separately because the
sum lies in ${\cal R}_m$. Moreover because $s_mB=0$ we can replace 
$s_m r_m$ by the anticommutator $\{s_m,r_m\}$ which counts the variables 
at level $m$
\beq
N_m=\{s_m,r_m\}=\sum_{m(\alpha)=m}
I_\alpha\frac{\partial}{\partial I_\alpha}+
q_\alpha\frac{\partial}{\partial q_\alpha}
\eeq
and maps ${\cal S}_m$ invertibly to itselves. Therefore $sP_m=0$ has 
only the trivial solution $P_m=0$.

It is, however, the form 
$\omega(C,F)=P_m(q_\alpha(C,F),I_\alpha(F)$ which has to satisfy 
$s\omega=0$, and not the polynomial $P_m(q_\alpha,I_\alpha)$. 
$s\omega=0$
can hold for nonvanishing $P_m$ if and only if the form degree
of $s\omega$ is larger than the dimension $D$ of
space time because the only additional algebraic identity which holds
for the composite variables $q_\alpha(C,F)$ and $I_\alpha(F)$ but not
for elementary anticommuting variables $q_\alpha$ and commuting
variables $I_\alpha$ comes from the fact that a product of more than D 
translation ghosts vanishes.
\beq
\prod_{i \in M}(I_{\alpha_i}(F))^{\beta_i}=0 \mbox{ if }\sum_{i \in M} 
2\beta_i m(\alpha_i)>D \quad \forall M \subset \{1,2,\dots ,r\}
\eeq
We obtain therefore the solutions $\omega$ if we take 
$B \in {\cal S}_m$ and restrict it in addition to be composed of
monomials with sufficiently many factors $I_\alpha$ such that the form
degree $D^\prime$ of $B$ lies above $D$ and the form degree of 
$\omega=r_mB$ starts below $D+1$. This restriction can
easily be formulated with the number operator
\beq
\label{nform}
N=\sum_\alpha 2m(\alpha)I_\alpha\frac{\partial}{\partial I_\alpha}
\eeq
which allows to split ${\cal S}_m$ into spaces ${\cal S}_{m,D^\prime}$
with definite and even degrees $D^\prime$.
\beq
{\cal S}_m=\sum_{D^\prime}{\cal S}_{m,D^\prime}\quad
P\in {\cal S}_{m,D^\prime}\Leftrightarrow P\in{\cal S}_m\wedge 
NP=D^\prime P
\eeq
Because each term in ${\cal S}_m$ contains at least one factor 
$I_\alpha$ with $m(\alpha)=m$ the degrees $D^\prime$ are not smaller
than $2m$. $D^\prime$ is restricted by
\beq
\label{dstrich}
D^\prime -2m \le D <D^\prime
\eeq
to obtain a nonvanishing solution 
\beq
\omega=\left (r_m P \right )_{|q_\alpha(C,F),I_\alpha(F)} 
\quad P\in {\cal S}_{m,D^\prime}
\eeq
which satisfies $s\omega=0$ because the translation ghost number of 
$s\omega$ is $D^\prime$  and larger than $D$.

If we want to obtain a solution $\omega$ with a definite ghost number
then we have to split the spaces ${\cal S}_{m,D^\prime}$ with 
the ghost counting operator $N_C$
\beq
N_C=\sum_\alpha \left (
2m(\alpha)I_\alpha\frac{\partial}{\partial I_\alpha} +
(2m(\alpha)-1)q_\alpha\frac{\partial}{\partial q_\alpha} \right )
\eeq
$N_C$ counts the total ghost number of translation ghosts, Lorentz 
ghosts and isospin ghosts and splits $ {\cal S}_{m,D^\prime} $ into
eigenspaces $ {\cal S}_{m,D^\prime,g}$ with total ghost number $g$
\beq
{\cal S}_{m,D^\prime}=\sum_g{\cal S}_{m,D^\prime,g}
\quad P\in {\cal S}_{m,D^\prime,g}\Leftrightarrow P\in {\cal
S}_{m,D^\prime}\wedge N_C P = g P
\eeq
The total ghost number of $\omega=r_mP$ is $g$ if 
$P\in {\cal S}_{m,D^\prime,g+1}$ because $r_m$ 
lowers the total ghost number by $1$.

We obtain the long sought solutions $\omega_D^g$ of the relative
cohomology (\ref{rel}) which for $g=0$ gives Lagrange densities of
invariant actions (\ref{relcoh}) and for $g=1$ gives anomaly candidates
(\ref{anomaly}) if we substitute in $\omega$ the ghosts $C$ 
by ghosts plus connection one forms $C+A$ and if we pick the part with 
$D$ differentials. 
Therefore the total ghost number $G$ of $P$ has to be chosen to be 
$G=g+D+1$ to obtain a solution $\omega$ which contributes to 
$\omega_D^g$. If the ghost variables $\hat{C}$(\ref{chat}) are used to 
express $\omega$ then $\omega_D^g$ is  simply obtained if all 
translation ghosts $C^m$ are replaced by $dx^m$ and the part with the 
volume element $d^Dx$ is taken.
\bea
\omega(C,F)&=&\left(r_m P\right)_{|q_\alpha(C,F),I_\alpha(F)} \quad P\in 
{\cal S}_{m,D^\prime,g+D+1}\\
\omega(C,F)&=&f(\hat{C}^m,\hat{C}^i,F^i)\\
\omega_D^g&=&f(dx^m,\hat{C}^i,F^i)_{|\mbox{\tiny D form part}}
\quad F^i={\scriptstyle \frac{1}{ 2}}dx^mdx^nF_{mn}{}^i
\eea
These  formulas end our general discussion of the BRS cohomology of 
gravitational Yang Mills theories.

Let us conclude by spelling out the general formula for $g=0$ and $g=1$.
If $g=0$ then $P$ can contain no factors $q_\alpha$ because the complete
ghost number $G\ge D^\prime$ is not smaller than the ghost number 
$D^\prime$ of translation ghosts. $D^\prime$ has to be larger than $D$ 
(\ref{dstrich}) and not larger than $G=g+D+1=D+1$ which leaves 
$D^\prime=D+1$ as only possibility. $D^\prime$ is even (\ref{nform}), 
therefore chiral contributions to Lagrange densities occur only in odd 
dimensions. 

If, for example $D=3$, then $P$ is an invariant 4 form. 

For $m=1$ such a form is given by $P=F_iF_j a^{ij}$ with $a^{ij}=a^{ji} 
\in \Real$ if the isospin group contains abelian factors with the 
corresponding abelian field strength $F_i$ and $i$ and $j$ enumerate
the abelian factors. $P$ lies in ${\cal S}_1$ because $P=s(q_iF_j
a^{ij})$. The form $\omega=r_1P=2q_iF_j a^{ij}$ yields the gauge
invariant abelian Chern Simons action in 3 dimension which is remarkable
because it cannot be constructed from tensor variables alone and because
it does not contain the metric.
 
To construct $\omega_3^0$ one has to express
$q(C)=C$ by $C=\hat{C}+C^mA_m$. Then one has to replace all translation 
ghosts by differentials $dx^m$ and to pick the volume form. One obtains
\beq
\omega_3^0{}_{abelian}=dx^mA_{mi}dx^kdx^lF_{klj} 
a^{ij}=\varepsilon^{klm}A_{mi}F_{kli}a^{ij}d^3x\ .
\eeq

For $m=2$ the form $P=trF^2$ of each nonabelian factor contributes to 
the nonabelian Chern Simons form. One has $I_1=tr F^2=sq_1$, so $P\in 
{\cal S}_2$ as required. $\omega$ is directly given by the Chern Simons 
form $q_1$(\ref{csimons})
\beq
\omega=tr(CF-\frac{1}{3}C^3)
\eeq
The corresponding Lagrange density is
\beq
\omega_3^0{}_{nonabelian}=tr(AF-\frac{1}{3}A^3)
\eeq

Chiral anomalies are obtained if one looks for solutions $\omega_D^1$
with ghost number $g=1$. This fixes $G=D+2$ and because $G$ is not less
than $D^\prime>D$ we have to consider the cases $D^\prime=D+1$ and 
$D^\prime=D+2$. 

The first case can occur in odd dimensions only, 
because $D^\prime$ is even, and only if the level $m$, the lowest degree 
occuring in P, is 1 because the missing total ghost number
$D+2-D^\prime$, which is not carried by $I_\alpha(F)$, has to be
contributed by one Chern Simons polynomial $q_\alpha$ with
$2m(\alpha)-1=1$, i.e. with $m(\alpha)=1$. Moreover $P\in {\cal S}_1$
and thefore has the form
\beq
P=\sum_{ij\ abelian}a^{ij}(I_\alpha)q_iI_j \quad a^{ij}=-a^{ji}
\eeq
where the sum runs over the abelian factors and the form degree 
contained in the antisymmetric $a^{ij}$ and in the abelian $I_j=F_j$ 
have to add up to $D+1$. In particular this anomaly can occur only if 
the gauge group contains at least two abelian factors because $a^{ij}$ 
is antisymmetric. In $D=3$ dimensions $a^{ij}$ is linear in abelian 
field strengths and one has
\beq
P=\sum_{ijk\ abelian}a^{ijk}q_iI_jI_k \quad a^{ijk}=a^{ikj} \quad 
\sum_{cyclic}a^{ijk}=0
\eeq
This leads to 
\beq
\omega=r_1P=\sum_{ijk\ abelian}b^{ijk}q_iq_jI_k
=\sum_{ijk\ abelian}b^{ijk}C_iC_jF_k  \quad b^{ijk}=-a^{ijk}+a^{jik}
\eeq
and the candidate anomaly is
\beq
\omega_3^1=2\sum_{ijk\ abelian}b^{ijk}\hat{C}_iA_jF_k=
\sum_{ijk\
abelian}b^{ijk}\hat{C}_iA_{m\,j}F_{rs\,k}\varepsilon^{mrs}d^3x\ .
\eeq

If one considers $g=1$ and $D=4$ then $D^\prime=6$ because it is bounded 
by $G=D+1+g=D+2$, larger than $D$ and even. This leaves 
$D^\prime=G$ as only possibility, so the total ghost number is 
carried by the translation ghosts contained in $P=P(I_\alpha)$ which is
a cubic polynomial in the field strength two forms $F$. Abelian two 
forms can occur in the combination
\beq
P=\sum_{ijk\ abelian}d^{ijk}F_iF_jF_k
\eeq
with completely symmetric coefficients $d^{ijk}$. One checks that these 
polynomials lie in ${\cal S}_1$. They lead to the abelian anomaly
\beq
\omega_4^1{}_{abelian}=\frac{3}{4}\sum_{ijk\ 
abelian}d^{ijk}\hat{C}_iF_{mn\ j} F_{rs\ k}\varepsilon^{mnrs}d^4x
\eeq
Abelian two forms $F_i$ can also occur in $P$  multiplied with 
$tr(F_k)^2$ where $i$ enumerates abelian factors and $k$ nonabelian
ones. The mixed anomaly which corresponds to
\beq
P=\sum_{ik}c^{ik}F_i tr(F_k)^2
\eeq
is very similar in form to the abelian anomaly
\beq
\omega_4^1{}_{mixed}=-\frac{1}{4}\sum_{ik}c^{ik}\hat{C}_i (\sum_I 
F_{mn}{}^I F_{rs}{}^I)_{k} \varepsilon^{mnrs}d^4x \ .
\eeq
The sum, however extends now over abelian factors enumerated by $i$ and 
nonabelian factors enumerated by $k$. Moreover we assumed that the 
basis, enumerated by $I$, of the simple Lie algebras is chosen such that
$trM_IM_J=-\delta_{IJ}$ holds for all $k$. Phrased in terms of $dA$ the 
mixed anomaly differs from the abelian one because the nonabelian
field strength contains also $A^2$ terms.

The last possibility to construct a polynomial $P$ with form 
degree $D^\prime=6$ is given by the Chern form $tr(F)^3$ itself. 
Such a Chern polynomial with $m=3$ exists for classical algebras only 
for the algebras $SU(n)$ for $n\ge 3$ and for $SO(6)$ (\ref{mlist}). In 
particular the Lorentz symmetry in $D=4$ dimensions is not anomalous.
The form $\omega$ which corresponds to the Chern form is the Chern 
Simons form
\beq
\omega(C,F)=tr \left ( CF^2-\frac{1}{2}C^3\,F + 
\frac{1}{10}C^5\right )\ .
\eeq
The nonabelian anomaly follows after the substitution $C\rightarrow C+A$ 
and after taking the volume form
\beq
\omega_4^1{}_{nonabelian}=tr(CF^2-\frac{1}{2}(CA^2F+ACAF+A^2CF)+\frac{1}
{2}CA^4)\ .
\eeq



\end{document}